\documentclass[11pt,letterpaper]{article}
\usepackage[margin=1in]{geometry}
\usepackage{fullpage}

\usepackage{float}
\usepackage{amsfonts,amsmath,amssymb,amsthm,graphicx}
\usepackage{amstext}
\usepackage{url}
\usepackage{subfig}
\usepackage[hidelinks,hypertexnames=false]{hyperref}
\usepackage[capitalize]{cleveref}
\usepackage{natbib}

\usepackage{thmtools}
\usepackage{thm-restate}

\usepackage{enumerate}

\usepackage{setspace}
\usepackage{multirow}
\usepackage{caption} 
\usepackage{enumitem}
\usepackage{listings}

\usepackage[linesnumbered,ruled]{algorithm2e} 
\SetKwComment{Comment}{/* }{ */}

\newtheorem{theorem}{Theorem}[section]
\newtheorem{corollary}[theorem]{Corollary}
\newtheorem{lemma}[theorem]{Lemma}
\newtheorem{proposition}[theorem]{Proposition}

\newtheorem{definition}[theorem]{Definition}
\newtheorem{assumption}[theorem]{Assumption}

\newtheorem{example}[theorem]{Example}

\newenvironment{proofof}[1]{\begin{proof}[Proof of #1]}{\end{proof}}

\newcommand{\given}{\,|\,}

\newcommand{\prob}[2][]{\text{\bf Pr}\ifthenelse{\not\equal{}{#1}}{_{#1}}{}\!\left[{\def\givenn{\middle|}#2}\right]}
\newcommand{\expect}[2][]{\text{\bf E}\ifthenelse{\not\equal{}{#1}}{_{#1}}{}\!\left[{\def\givenn{\middle|}#2}\right]}

\newcommand{\tparen}{\big}
\newcommand{\tprob}[2][]{\text{\bf Pr}\ifthenelse{\not\equal{}{#1}}{_{#1}}{}\tparen[{\def\given{\tparen|}#2}\tparen]}
\newcommand{\texpect}[2][]{\text{\bf E}\ifthenelse{\not\equal{}{#1}}{_{#1}}{}\tparen[{\def\given{\tparen|}#2}\tparen]}

\newcommand{\sprob}[2][]{\text{\bf Pr}\ifthenelse{\not\equal{}{#1}}{_{#1}}{}[#2]}
\newcommand{\sexpect}[2][]{\text{\bf E}\ifthenelse{\not\equal{}{#1}}{_{#1}}{}[#2]}

\newcommand{\eps}{\epsilon}

\renewcommand{\P}{\mbox{\sf P}}
\newcommand{\NP}{\mbox{\sf NP}}

\usepackage{xcolor}

\usepackage{color-edits}
\usepackage{thm-restate}

\begin{document}

\title{Welfare in Envy-free Contracts: When Agents Have Sparse Interests} 
\author{
Junjie Chen\thanks{School of Computer Science and Engineering, Southeast University. Email: \texttt{junjie.chen@seu.edu.cn}}
\and Cong Chen\thanks{School of Management, Guangzhou University. Email: \texttt{chencong@gzhu.edu.cn}}
\and Takuro Yamashita\thanks{ISER, The University of Osaka. Email: \texttt{yamashita@iser.osaka-u.ac.jp} }
}
\date{}

\begin{titlepage}
	\clearpage\maketitle
	\thispagestyle{empty}

\begin{abstract}
We study welfare maximization under envy-free contracts, where a principal assigns tasks and offers success-contingent payments to heterogeneous agents with hidden effort. For unit-demand agents, we establish a sharp complexity boundary: the problem is polynomial-time solvable when each agent is interested in at most two tasks, but NP-hard with at most three interests. Our algorithm analyzes tree-structured dependencies among forced assignments and reduces optimization to maximum-weight independent set in a perfect graph. When agents may receive multiple tasks, no polynomial-time constant-factor approximation exists unless $\mathrm{P}=\mathrm{NP}$. Finally, with homogeneous success probabilities and rewards, we establish a welfare- and revenue-preserving equivalence between unit-demand contract design and envy-free pricing. This yields polynomial-time welfare maximization without restricting interests, while revenue maximization remains NP-hard.
\end{abstract}

\end{titlepage}

\section{Introduction}\label{section_intro}

Contract design studies how a principal can use monetary payments to incentivize agents to take costly actions that are not directly observable. This problem, commonly known as moral hazard, arises when a principal delegates work but can observe only its outcome. A service provider, for example, may privately decide how much effort to invest in a task, while the principal observes only whether the task succeeds. By conditioning payments on observable outcomes, a contract aligns the agent's incentives with the principal's objectives. The algorithmic study of contracts investigates the computational complexity of optimal incentive design and the performance of simple contract classes \citep{dutting2024algorithmic}. Linear contracts, which pay a fixed fraction of the realized reward, are a central example of such simple incentives \citep{dutting2019simple}.

With multiple agents, contract design also raises questions of fairness. An agent may be willing to perform its assigned task while preferring the task and compensation offered to another agent. Such comparisons depend on both the work being assigned and the payment it carries: agents can have different costs and success probabilities for the same task. \citet{castiglioni2025algorithmic} introduce a fair contract design framework based on \emph{envy-freeness} (EF). Each agent must weakly prefer its own task--contract bundle to every other agent's bundle, evaluated using its own costs, success probabilities, and optimal effort choices. This requirement connects contract design to fair allocation and calls for coordinating assignments and compensation across agents.

In this paper, we study \emph{welfare maximization in envy-free contracts}. A principal allocates tasks to heterogeneous agents and chooses a task-level linear contract for each task. If agent $i$ exerts effort on task $j$, it incurs cost $c_{i,j}$ and succeeds with probability $p_{i,j}$, generating reward $r_j$ upon success. A contract $\alpha_j\in[0,1]$ pays the agent a fraction $\alpha_j$ of that reward. Every assigned agent must be willing to exert effort, and the allocation and contracts must satisfy envy-freeness. We allow tasks to remain unassigned. Our main focus is on \emph{unit-demand agents}, each of whom can be assigned at most one task.
The welfare contribution of assigning task $j$ to agent $i$ is the expected reward net of the effort cost, $W_{i,j}=p_{i,j}r_j-c_{i,j}$. For a fixed allocation, this objective does not depend on the contracts. Nevertheless, contracts determine whether the allocation can satisfy both effort incentives and envy-freeness. Without envy-freeness, the unit-demand problem can be solved by a maximum-weight bipartite matching. Under envy-freeness, the payment needed to induce effort on one task can create envy from other agents and force additional assignments. Consequently, maximizing welfare requires understanding which matchings can be supported by fair contracts.

We investigate whether the problem becomes tractable when agents have \emph{sparse interests}. Agent $i$ is interested in task $j$ if $p_{i,j}r_j-c_{i,j}>0$, meaning that the task can give the agent positive utility under some admissible contract. An instance is $k$-interest if every agent is interested in at most $k$ tasks. This restriction captures specialization: each provider can profitably undertake only a small number of the available tasks. It limits the alternatives relevant to an individual agent's envy comparisons, while allowing a task to attract arbitrarily many agents. Our central question is how this sparsity affects the complexity of welfare maximization under exact envy-freeness.

\subsection{Our Results and Proof Strategy}\label{subsec:results_strategy}

We establish a sharp complexity boundary between two and three interests for unit-demand agents. We also prove a general inapproximability result and identify a homogeneous setting that is equivalent to the classical envy-free pricing.

\paragraph{A polynomial-time algorithm for two interests.}
Our main positive result shows that sparse interests can make exact welfare maximization tractable.

\vspace{2mm}
\noindent\textbf{Theorem 1. (\cref{polynotimaitmelaoghtm})} \emph{When agents are unit-demand and each agent is interested in at most two tasks, a welfare-maximizing envy-free allocation and supporting contracts can be computed in polynomial time.}
\vspace{2mm}

Our approach draws inspiration from \citet{chen2010envy}, who reduce revenue-maximizing envy-free pricing in two-interest, unit-demand, unit-supply markets to maximum-weight independent set in a perfect graph. Both approaches encode the compatibility of allocation choices in a graph. However, extending this approach to fair contracts introduces two additional challenges. First, agents' effort choices are hidden, so the principal must jointly choose assignments and contracts that satisfy both effort incentives and envy-freeness. Second, an item's price enters every buyer's utility with the same coefficient, whereas a task's contract share enters an agent's utility with the agent-specific coefficient $p_{i,j}r_j$. These additional incentive constraints and heterogeneous coefficients prevent a direct application of the pricing algorithm and require new structural arguments to characterize supporting contracts and compatibility. 

The key technical observation is that a single assignment can trigger many further assignments. If the payment on an assigned task gives an unmatched agent positive utility, envy-freeness requires that agent to receive a task as well. With at most two interests, the agent has at most one alternative task, so its assignment is forced. This assignment may create envy from another agent, generating dependencies involving many agents and tasks even in sparse instances.
We capture these dependencies through the delicately constructed \emph{minimum matching} induced by a base assignment. 
This matching consists of assignments that must appear in every feasible outcome containing the base assignment. We can show that the induced minimum matching structure essentially forms a \emph{tree}. Unlike \citet{chen2010envy} deriving chains anchored at buyer--item pairs where the buyer pays its \emph{highest price} (i.e., full valuation), the minimum matching tree is found 
through coordinatewise \emph{least contracts} satisfying effort and envy constraints among the currently matched agents. 

The next challenge is to characterize how minimum matchings interact. \citet{chen2010envy} derive conditional transitivity and symmetry relations through chains-based analysis. Our analysis is tree-based: we track how contract inequalities propagate along paths and control the assignments forced on off-path branches. Specifically, we examine how base assignments and their opposite assignments occur within other minimum matchings, deriving nesting relations for compatible matchings and structural characterizations of different types of conflicts.
Moreover, while their selected price-setter components partition the sold items, our compatible minimum matchings may overlap. We prove that these overlaps are nested and that compatible matchings can be combined into a feasible outcome using coordinatewise maxima of their supporting contracts.

We then construct a conflict graph with one vertex for each feasible base assignment. Two vertices are adjacent when their induced minimum matchings contain conflicting assignments. Assigning each vertex the welfare of its base assignment reduces welfare maximization to maximum-weight independent set. Using the structural relations established above, we exclude odd holes and odd antiholes, proving that the graph is perfect by the Strong Perfect Graph Theorem \citep{chudnovsky2006strong}. Polynomial-time optimization over perfect graphs then yields our algorithm \citep{grotschel1981ellipsoid}.

\paragraph{NP-hardness for three interests.} Despite the additional incentive constraints and heterogeneous coefficients in fair contract design, the same two-versus-three interest boundary emerges as \citet{chen2010envy}: welfare maximization is polynomial-time solvable for unit-demand agents with at most two interests and NP-hard with at most three.

\vspace{2mm}
\noindent\textbf{Theorem 2. (\cref{thm:welfare-three-interests})}  \emph{When agents are unit-demand, welfare maximization under effort and envy-free constraints is NP-hard even if every agent is interested in at most three tasks.}
\vspace{2mm}

The proof reduces from $3$-SAT. A variable agent chooses between two root tasks, encoding a truth assignment. Each clause has three literal agents, and a literal agent is interested in exactly three tasks: its private task, a common clause task, and the variable task corresponding to the literal being false. A dummy agent forces the common clause task to carry a zero share when it is assigned to a literal agent. Envy-freeness then permits that literal agent to receive the clause task only when its literal is true. A carefully chosen welfare threshold forces the variable choices and a literal assignment for every clause. Reaching that threshold is thus equivalent to satisfying the formula, implying the $\NP$-hardness results.

\paragraph{Inapproximability in the general setting.}
We also consider welfare-maximizing in general settings, where agents may receive multiple tasks, with no bound on their interests.

\vspace{2mm}
\noindent\textbf{Theorem 3. (\cref{theorem_harndssresults})}  \emph{Unless $\mathrm{P}=\mathrm{NP}$, no polynomial-time algorithm achieves any constant-factor approximation to the optimal welfare of an envy-free contract in the general setting.}
\vspace{2mm}

The proof uses a gap reduction from Independent Set. Vertex tasks generate substantial welfare when assigned to a designated agent. Agents associated with graph edges impose envy constraints that prevent the designated agent from receiving both endpoint tasks of an edge. The attainable welfare consequently reflects the size of an independent set, transferring the approximation gap to fair contract design.

\paragraph{A connection to envy-free pricing.}
Finally, we isolate the role of heterogeneous production parameters and relate fair contract design to the classical envy-free pricing problem.

\vspace{2mm}
\noindent\textbf{Theorem 4. (\cref{reudction_eachotuerh})}  \emph{For unit-demand agents with a common success probability and a common task reward, fair contract design and unit-demand, unit-supply envy-free pricing reduce to each other in polynomial time, preserving welfare and revenue.}
\vspace{2mm}

Writing the common probability and reward as $p$ and $r$, an agent's utility from an assigned task becomes
$\alpha_jpr-c_{i,j} = (pr-c_{i,j})-(1-\alpha_j)pr.$
Intuitively, for pairs with nonnegative net surplus, the first term acts as the agent's valuation, while the principal's retained expected reward acts as an item price. This identity supports reductions in both directions. Through the equivalence, welfare maximization is polynomial-time solvable in the homogeneous setting using a welfare-maximizing matching and supporting competitive prices, whereas revenue maximization remains NP-hard. Thus, homogeneous success probabilities and rewards yield tractability even without sparse interests.

\subsection{Related Work}\label{subsec:related_work}

\paragraph{Algorithmic contract theory.}
Contract theory has a rich literature in microeconomics dating back several decades \citep[e.g.,][]{holmstrom1980theory,myerson1982optimal,carroll2015robustness,gottlieb2015simple,chade2019disentangling}. Its algorithmic study examines computational tractability, approximation, and learning, including combinatorial models~\citep{babaioff2006combinatorial}. For example, \citet{dutting2019simple} analyze approximation guarantees for linear contracts, while \citet{dutting2021complexity} investigate the complexity of computing optimal contracts. Further work develops approximation algorithms~\citep{deo2024supermodular} and studies learning contracts from data~\citep{bacchiocchi2023learning,chen2024bounded,han2024learning}; see \citet{dutting2024algorithmic} for a survey.

Another line of research incorporates adverse selection by allowing agents to have private types. This literature studies single-parameter contracts~\citep{alon2021contracts,alon2022bayesian}, multi-parameter contracts~\citep{castiglioni2021bayesian,castiglioni2022designing,guruganesh2021contracts,guruganesh2023power}, and approximation-preserving reductions between these settings~\citep{castiglioni2025reduction}. Other directions include ambiguous contracts~\citep{dutting2023ambiguous} and the joint design of information and contracts~\citep{castiglioni2025hiring}.

Multi-agent contract design studies teams whose hidden actions jointly determine a project's outcome~\citep{dutting2023multi}. Extensions allow multiple actions per agent~\citep{cacciamani2024multi}, combinatorial actions~\citep{duetting2025multi}, and multiple projects~\citep{alon2025multi}. In single-agent combinatorial contracts, the agent chooses a subset of actions~\citep{dutting2022combinatorial}; related work investigates settings beyond gross substitutes~\citep{dutting2024combinatorial}, sequential actions~\citep{ezra2024contract}, and inspections~\citep{ezra2024contracts}. Our work focuses on allocating distinct tasks and choosing task-level contracts to maximize welfare subject to effort incentives and envy-freeness.

\paragraph{Fair contract design.} The algorithmic study of fair contract design was initiated by \citet{castiglioni2025algorithmic}, whose task-level contracting model is closest to ours. They introduce envy-free contracts, establish existence under full allocation, and study revenue optimization, relaxed fairness notions, and the price of fairness, providing algorithms for settings with a constant number of agents or tasks. Building on this framework, we permit partial allocation, optimize welfare, and identify sparse interests as a source of tractability. Recently, \citet{castiglioni2026envy} introduce agent-specific subsidies to study the fairness--revenue tradeoff under this framework.

Subsequent work explores other approaches to fairness in contracting. \citet{castiglioni2025fair} study fairness within a team working on a joint project, accounting for how swapping contracts can change agents' effort choices and the common outcome. \citet{feldman2026equal} study equal-pay team contracts, which satisfy the fairness notion of \citet{castiglioni2025fair}, and establish approximation and hardness results together with tight bounds on the revenue loss from requiring equal payments. Our model concerns separate tasks, with envy-freeness comparing the task--contract bundles assigned to different agents.

\paragraph{Envy-free pricing.}
Envy-free pricing studies revenue-maximizing prices and allocations that give each buyer a utility-maximizing bundle \citep{guruswami2005profit}. Subsequent work develops approximation algorithms and simple pricing schemes \citep{balcan2006approximation,balcan2008item}, establishes conditional inapproximability \citep{briest2008uniform,chalermsook2012improved}, and investigates supply constraints \citep{cheung2008approximation,im2010envy} and large markets \citep{anshelevich2017envy}. Particularly relevant are the perfect-graph approaches of \citet{chen2008optimal} for metric substitutability and \citet{chen2010envy} for unit-demand, unit-supply markets with at most two interests per buyer; the latter also establish NP-hardness with three interests. Our analogous boundary concerns welfare maximization under hidden effort and heterogeneous success probabilities. In standard unit-demand assignment markets, welfare-maximizing matchings admit supporting competitive prices \citep{blumrosen2007combinatorial}. Our equivalence transfers this tractability to contracts with homogeneous success probabilities and rewards, and clarifies why pricing results alone do not resolve welfare maximization in the heterogeneous contract setting.


\section{Model}

We consider the standard envy-free contract model as in \citep{castiglioni2025algorithmic}, where
a principal is allocating a set of tasks $\mathcal{T}$ to a set of agents $\mathcal{A}$.  Let $|\mathcal{A}| = n$ and $|\mathcal{T}|=m$.
Each task has two outcomes: either an agent successfully completes the task or the task fails.
For each agent $i \in \mathcal{A}$ and a task $j\in \mathcal{T}$, if agent $i$ exerts effort on task $j$ at a cost $c_{i,j} \ge 0$, it will succeed in task $j$ with a success probability $p_{i,j} \in [0,1]$. Upon the success of task $j$, the principal receives reward $r_j\ge 0$; otherwise, the reward is zero. If agent $i$ does not exert effort, it incurs no cost and the task is deemed to fail. 

The agents' choices of effort are private and hidden from the principal, while the principal can only observe the outcome of the task. Hence, the principal incentivizes the agent through payments conditional on realized outcomes. The principal designs a task-level {\it linear} contract $\alpha_j \in [0, 1]$ for each task $j\in \mathcal{T}$, and simultaneously allocate task  $j$ to some agent.  Following~\citep{castiglioni2025algorithmic}, we restrict our attention to settings where each task can only be allocated to at most one agent, while each agent may be allocated a set of tasks, denoted as $S_i \subseteq\mathcal{T}$. Note that, different from~\citep{castiglioni2025algorithmic},  we do not assume a full allocation of tasks, i.e., we only require  $\bigcup_{i\in \mathcal{A}} S_i \subseteq \mathcal{T}$.

Given a task $j$ and its associated contract $\alpha_j$, the agent $i$ will choose to exert effort on task $j$ if it receives non-negative utility, i.e., $\alpha_j p_{i,j} r_j -c_{i,j} \ge 0$. As in~\citep{castiglioni2025algorithmic}, a set of tasks $S_i$ can be assigned to agent $i$ if agent $i$ is willing to exert effort on every task in $S_i$. That is, the following effort constraints must hold
\begin{equation}\label{eff_constraint}
     \alpha_j p_{i,j} r_j -c_{i,j} \ge 0 \quad \forall i\in \mathcal{A}, j\in S_i. \tag{Effort constraints}
\end{equation}

Additionally, when designing the allocation of tasks, the envy-free contracts model requires that each agent prefers its own assigned set of tasks over any other agent's set of tasks. In other words, each agent $i\in \mathcal{A}$ obtains weakly higher utility from set $S_i$ than from any other set $S_j, j\neq i$ by exerting optimal efforts. We formulate it as an envy-free (EF) constraint,
\begin{equation}\label{ef_cons_eq}
    \sum_{k\in S_i} (\alpha_k p_{i,k} r_k -c_{i,k})  \ge \sum_{k \in S_{j}} \max\{ \alpha_k p_{i,k} r_k -c_{i,k}, 0\}, \quad \forall i \in \mathcal{A}, j \in \mathcal{A} \tag{EF constraints}
\end{equation}

\paragraph{Welfare-maximizing principal.} In this paper, the principal cares about the welfare derived from the design. 
Define the welfare obtained from agent $i$ performing task $j$ as 
\[
W_{i,j} = p_{i,j}r_j -c_{i,j}
\]
Given any contracts and allocation $(S, \alpha)$, the principal's welfare is defined as
\[
\texttt{Wel}(S, \alpha) = \sum_{i \in \mathcal{A}} \sum_{j \in S_i} W_{i,j}.
\]
 Hence, subject to envy-free constraints and the effort constraints for allocation, the principal's welfare-maximizing problem is formulated as follows,
\begin{equation}\label{program-welfare}\tag{FCW}
   \begin{aligned}
    \max_{S, \alpha } \quad & \texttt{Wel}(S, \alpha) \\
    \text{subject to.} \quad & \text{\eqref{eff_constraint}, \eqref{ef_cons_eq}}
\end{aligned} 
\end{equation}

\paragraph{Agent's demand and interests.} In this paper, we are particularly interested in the unit-demand agent, while \cref{hardness_general_case_sec} concerns settings where agents have a general demand.
\begin{assumption} \label{assump_epdn_c2}
(Unit-demand agent)
    Each agent can be assigned at most one task.
\end{assumption}
Next, we define agents' interests. An agent is said to be interested in one task if and only if it can {\it potentially} receive a positive utility.
\begin{definition}
    An agent $i\in \mathcal{A}$ is interested in a task $j\in \mathcal{T}$ if $p_{i,j}r_j - c_{i,j} > 0$. We call an instance \emph{$k$-interest} if each agent is interested in at most $k$ tasks.
\end{definition}

Under Assumption~\ref{assump_epdn_c2}, we can view the welfare-maximizing problem \eqref{program-welfare} as a matching problem. Specifically, the positive-welfare assignments form a bipartite graph $\mathcal{G}=(\mathcal{A}\cup\mathcal{T}, \mathcal{E})$. There are $n$ nodes on the left corresponding to the set of agents, and there are $m$ nodes on the right corresponding to the set of tasks. If agent $i$ is interested in task $j$, an edge $e=(i,j) \in \mathcal{E}$ between node $i$ on the left and node $j$ on the right exists.
Hence, the welfare-maximizing problem is equivalent to finding a maximum-weight bipartite matching where \eqref{eff_constraint} and \eqref{ef_cons_eq} need to be satisfied. 


\section{Hardness of General Cases}\label{hardness_general_case_sec}
In this section, we first present hardness results for the \eqref{program-welfare} problem in general settings: agents have general demands, and each agent is interested in any number of tasks. We show that finding the optimal fair allocation for \eqref{program-welfare} is computationally challenging.
In particular, we show that even finding an approximately optimal envy-free contract is $\NP$-hard for any constant approximation ratio

\begin{theorem}\label{theorem_harndssresults}
    For any constant $c\ge 1$, unless $\P=\NP$, there does not exist a polynomial-time algorithm that computes an EF contract with welfare at least a $\frac{1}{c}$ fraction of the optimum.
\end{theorem}

Our reduction is inspired by \cite{castiglioni2025algorithmic}, but it reduces from a different problem, which is deciding whether there exists an independent set involving at least
some large fraction of vertices or all the independent sets have at most some small fraction of vertices ~\citep{zuckerman2006linear}.  Formally, for any $\delta >0$, given a graph $G=(V, E)$, it is NP-hard to decide between the two following cases:
\begin{itemize}
    \item \textsf{Yes case}: there exists an independent set of size at least $|V|^{1-\delta}$
    \item \textsf{No case}: all independent sets are of size smaller than $|V|^{\delta}$.
\end{itemize}

\begin{proofof}{\cref{theorem_harndssresults}}
   Given a graph $G=(V, E)$, let $\gamma = |V|$, $\beta = |E|$,  and define $\eps = \frac{1}{\gamma\beta}$.
    \paragraph{Construction.} For each edge $e\in E$, there exists an agent $b_{e}$. Additionally, we have one ``effective" agent $b^*$.
	Then, we define two sets of tasks: 
	\begin{itemize}
		\item For each vertex $v\in V$, we have a task $d_v$ with reward $r_{d_v}=1$. The probabilities of success and costs are constructed as follows.
		\begin{itemize}
			\item Let $p_{b^*,d_v}=1$ and $c_{b^*,d_v}=1/2$.
			\item If node $v\in e$, i.e.,  the edge $e$ connects $v$ to another node, then let  $p_{b_{e},d_v}=\eps$ and $c_{b_{e},d_v}=0$.
			\item Let $p_{b,d_v}=0$ and $c_{b,d_v}=0$  for all other agents $b$.
		\end{itemize}
		\item For each edge $e\in E$, we define a task $ d_{e}$ with $r_{d_{e}}=\eps/2$.  The probabilities of success and costs are defined as follows.
		\begin{itemize}
			\item For agent $b_{e}$, let $p_{b_{e}, d_{e}}=1$ and $c_{b_{e}, d_{e}}=0$.
			\item Let $p_{b,d_{e}}=c_{b,d_{e}}=0$ for all other agents $b$.
		\end{itemize}
	\end{itemize}
    
\paragraph{\textsf{Yes case}.} In this case, there exists one independent set $I$ of size at least $|I|\ge |V|^{1-\delta}$. We assign the set of tasks corresponding to the independent set to ``effective" agent $b^*$, and set the contract $\alpha_j = \frac{1}{2}$ for $j\in I$. Moreover, assign task $d_e$ to each agent $b_e$ for $e\in E$, and assign a contract $\alpha_{d_e} = 1$. All other tasks are not allocated. Clearly, there is no envy between agents $b_e$ and $b_{e'}$ for $e, e'\in E$. Moreover, agent $b^*$ does not envy agent $b_e$ for every  $e\in E$. Since $I$ is an independent set, for each edge $e\in E$, at most one task $d_v$ in edge $e$ is assigned to agent $b^*$. The utility that agent $b_e$  get from set of tasks assigned to $b^*$ is $\alpha_{d_v} p_{b_{e},d_v} r_{d_v} - c_{b_{e},d_v} = \frac{\eps}{2}$. Therefore, agent $b_e$ does not envy agent $b^*$. The welfare is generated from assignments of set of tasks $I$, and the tasks $d_e$ for all $e \in E$. This gives welfare $\frac{1}{2}\gamma^{1-\delta} + \frac{\beta\eps}{2} \ge \frac{1}{2}\gamma^{1-\delta}$.

\paragraph{\textsf{No case}.}

Suppose all independent sets has size $\gamma^{\delta}$.
Consider an allocation $S$ and a contract $\alpha$.
Let $V^*$ be the vertex related to the tasks assigned to agent $b^*$, i.e., $V^*=\{v| v\in S_{b^*}\} \subseteq V$. 
Notice that by effort constraints, $\alpha_{d_v}\ge \frac{1}{2}$ for all $v \in V^*$.

We show that $V^*\le \gamma^{\delta}$. Indeed, otherwise $V^*$ is not an independent set.
If $V^*$ is not an independent set, it must be the case that there exists an edge $e=(v,v')$ with $v,v' \in V^*$. Hence, the envy free constraints of agent $b_{e}$ implies that such agent receive payment at least $\epsilon$. 
 It is easy to see that this is not possible, as agent $b_e$ can receive payment at most $r_{d_e}=\frac{\eps}{2}$ from task $d_e$, reaching a contradiction.
Hence, we can conclude that $V^*\le \gamma^{\delta}$
he total welfare is at most 
\[
\frac{1}{2}|V^*| + \eps (|V|-|V^*|) + \frac{\eps}{2} |E| \le \frac{\gamma^\delta}{2} + \frac{1}{\beta} + \frac{1}{2\gamma} 
\]

\paragraph{Gap.} The gap between two cases is $\frac{\frac{\gamma^\delta}{2} + \frac{1}{\beta} + \frac{1}{2\gamma} }{\frac{1}{2}\gamma^{1-\delta}} \le \frac{\frac{\gamma^\delta}{2} + 2}{\frac{1}{2}\gamma^{1-\delta}} =4 \gamma^{\delta-1} + \gamma^{2\delta-1}$ which can be arbitrarily small as $\gamma$ is large and $\delta\to 0$. This concludes the proof.
\end{proofof}

\section{A Polynomial-Time Algorithm for $2$-interests Instances}\label{sec:two_interests}

From this section on, we start to focus on cases of unit-demand agents. Specifically, in this section, we show that, if the agents are unit-demand and interested in at most $2$ tasks, the welfare-maximizing problem can be solved in polynomial time by reducing it to the problem of finding a maximum-weight independent set in a perfect graph.

\begin{theorem}\label{polynotimaitmelaoghtm}
    When agents are unit-demand and interested in at most $2$ tasks, there exists a polynomial-time algorithm for computing a welfare-maximizing solution to the problem~\eqref{program-welfare}.
\end{theorem}

Throughout this section, we maintain the following notations. For a matching $M$, write $\mathcal{A}(M)$ and $\mathcal{T}(M)$ for its matched agents and assigned tasks. We use
\[
 U_{i,j}(\alpha_j)=\alpha_j p_{i,j}r_j-c_{i,j}
\]
for the utility from exerting effort. A \emph{feasible matching profile} $(M,\alpha^M)$ satisfies the effort and EF constraints for every agent, including unmatched agents. We set $\alpha^M_j=0$ on unassigned tasks. In particular, an unmatched agent $i$ must satisfy $U_{i,j}(\alpha^M_j)\le0$ for every $j\in\mathcal{T}(M)$. Moreover, for $e=(i,j)$, let $\operatorname{opp}(e)=(i,t)$ if $t$ is the owner's other interested task; If $t$ does not exist, then $\operatorname{opp}(e)=\emptyset$.

The key step of the proof is to construct a conflict graph $\mathcal{C}$, where each node $e$ in the graph $\mathcal{C}$ corresponds to the matching induced by the edge $e\in \mathcal{E}$. Our first step is to construct such induced matchings.

\paragraph{The matching $\mathcal{B}_e$ induced by edge $e \in \mathcal{E}$.}  Consider agents $i, j$ who both are interested in task $k$. A crucial observation is that if the principal assigns task $k$ to agent $i$, it may need to assign some other task $t$ to agent $j$ so as to satisfy the EF constraints. We illustrate it using  \cref{EFassigningmustassign}.

\begin{example}\label{EFassigningmustassign}
    There are two task with rewards $r_1=r_2 =1$. There are two agents. For agent $1$, its success probabilities and costs regarding two tasks are $c_{1, 1} = 0.1, p_{1, 1} = 0.2$ and $c_{1, 2} = 0.2, p_{1, 2} = 0.1$, respectively. For agent $2$, its success probabilities and costs are $c_{2, 1} = 0.1, p_{2, 1} = 0.5$ and $c_{2, 2} = 0.1, p_{2, 2} = 1$. If the principal assigns task $1$ to agent $1$, the contract is at least $\alpha_1 \ge \frac{c_1}{p_{1, 1}r_1} = \frac{1}{2}$. This causes envy from agent $2$. To satisfy the EF constraints, the principal must also assign task $2$ to agent $2$.
\end{example}
Note that when the principal further assigns task $t$ to agent $j$ to satisfy EF constraints, it may also cause envy from a third agent $l\neq i ,j$ who is interested in task $t$. Therefore, the EF constraint further forces task assignment to agent $l$. If this assignment process continues and finally terminates until all the agents $\mathcal{A}$ satisfy EF and IR constraints, we say that this set of assignments is induced by $e =(i, k)$, where $e=(i,k) \in \mathcal{G}$ represents assigning task $k$ to agent $i$. We denote this set of assignment as $\mathcal{B}_e$, and clearly, $\mathcal{B}_e$ is a matching. 

\begin{definition} A minimum matching $\mathcal{B}_e$ induced by edge $e = (i, j) \in \mathcal{E}$ is 
    a set of assignments that {\bf must be} included because of  assigning task $j$ to agent $i$ {\bf at first}. In other words, for every feasible matching $M$ that contains $e$, it should hold that $\mathcal{B}_e \subseteq M$.
\end{definition}

We emphasize that these assignments are included by a \emph{\bf{must}}. There are cases where assignments are not necessarily added, illustrated in \cref{addassignment}.
\begin{example}\label{addassignment}
Continue with \cref{EFassigningmustassign}. If the principal assigns the task $2$ to agent $2$ under contract $\alpha_2 = 0.6$, it will cause agent $1$'s envy, and EF constraints force to assign task $1$ to agent $1$. However, the assignment of task $1$ to agent $1$ is {\bf not} a must. Indeed, if  $\alpha_2 = 0.1$, agent $2$ still exerts efforts on task $2$, but it will not cause envy from agent $1$.
\end{example}

Next, we show how to find the matching $\mathcal{B}_e$. 
Given a fixed matching $M$ (not necessarily feasible), let $\mathcal{X}^M$ be the set of contracts that satisfy the following constraints, 
\begin{equation}\label{eq:owner_constraints}
 \begin{aligned}
 0\le\alpha_j\le1 &\qquad(j\in\mathcal{T}(M)),\\
 \alpha_j = 0 &\qquad(j\notin\mathcal{T}(M)),\\
 U_{i,j}(\alpha_j)\ge0 &\qquad((i,j)\in M),\\
 U_{i,j}(\alpha_j)\ge U_{i,t}(\alpha_t)
 &\qquad((i,j)\in M,\ t\in\mathcal{T}(M)).
 \end{aligned}
\end{equation}
That is for every $\alpha^M \in \mathcal{X}^M$, we only require that the \eqref{eff_constraint} and \eqref{ef_cons_eq} are satisfied for the set of agents $\mathcal{A}(M)$ and tasks $\mathcal{T}(M)$.

\begin{lemma}\label{minimum_payment}
If $\mathcal{X}^M$ is nonempty, it has a unique coordinatewise least element $\underline\alpha^M$ such that $\underline\alpha^M_j = \min_{\alpha \in \mathcal{X}^M} \alpha_j$ for all $j\in \mathcal T$. Moreover, $(M, \underline\alpha^M)$ forms a feasible matching profile, i.e., $\underline\alpha^M \in \mathcal{X}^M$.
\end{lemma}
\begin{proof}
    Note that by the constraints in \eqref{eq:owner_constraints}, we know that the feasible contract set $\mathcal{X}^M$ is bounded and closed, and thus a compact set. This means that the minimum $\underline\alpha^M_j = \min_{\alpha \in \mathcal{X}^M} \alpha_j$ exists for every task $j$.  
    Hence, $\underline\alpha^M$ exists.

Next, we show that $(M, \underline\alpha^M)$ forms a feasible matching. Clearly, the IR constraints continue to hold for agents in $\mathcal{A}(M)$ by the definition of $\underline\alpha^M$. The remaining proof is to show that the EF constraints continue to hold. Pick an arbitrary contract $\alpha^M \in \mathcal{X}^M$.  Consider any two agents  $i,j \in \mathcal{A}(M)$, and suppose task $s$ is assigned to agent $i$ and task $t$ is assigned to agent $j$. By the definition of $\underline\alpha^M$, we know that there must exist some contract $\alpha^M \in \mathcal{X}^M$ such that $\underline\alpha^M_s = \alpha^M_s$.
We show agent $i$ does not envy agent $j$ under contract $z$, since
\[
r_s p_{i,s}\underline\alpha^M_s - c_{i,s} = r_s p_{i,s}\alpha^M_s - c_{i,s} \ge \max\{r_t p_{i,t}\alpha^M_t - c_{i,t}, 0\} \ge \max\{r_t p_{i,t} \underline\alpha^M_t - c_{i,t}, 0\}
\]
where the first inequality is that $(M, \alpha^M)$ satisfy \eqref{eq:owner_constraints}. This completes the proof.
\end{proof}

A straightforward corollary of \cref{minimum_payment} is that $\underline\alpha^M = \arg\min_{\alpha \in \mathcal{X}^M} \sum_{j} \alpha_j$ gives the minimum sum of payments, and it is unique. Moreover, such payment minimum property and uniqueness extend to cases where $M$ is a feasible matching, i.e., both the constraints in \eqref{eq:owner_constraints} and $U_{i,j}(\alpha_t) \le 0, i\notin \mathcal{A}(M), t\in \mathcal{T}(M)$ hold. Let $\mathcal{X}^M$ be the set of contracts such that $M$ is a feasible matching. 

\begin{corollary}
    \label{minimum_payment_extend}
If $\mathcal{X}^M$ is nonempty where $M$ is a feasible matching, it has a unique coordinatewise least element $\underline\alpha^M$ such that $\underline\alpha^M_j = \min_{\alpha \in \mathcal{X}^M} \alpha_j$ for all $j\in \mathcal T$. Moreover, $(M, \underline\alpha^M)$ forms a feasible matching profile, i.e., $\underline\alpha^M \in \mathcal{X}^M$.
\end{corollary}
\begin{proof}
    We know that $\mathcal{X}^M$ is compact. Hence, $\underline\alpha^M$ exists. To show the claim, in addition to the proof of \cref{minimum_payment}, we only need to prove that \eqref{ef_cons_eq} holds for unassigned agents.
    Consider unassigned agent $i\in \mathcal{A}\setminus \mathcal{A}(M)$ and agent $j \in \mathcal{A}(M)$, and suppose task $k$ is assigned to agent $j$. Pick an arbitrary contract $\alpha^M \in \mathcal{X}^M$. Agent $i$ does not envy agent $j$ under contract $\underline\alpha^M$, since
\[
0 \ge \max\{r_k p_{i,k} \alpha^M_{k} - c_{i, k}, 0\}\ge \max\{r_k p_{i,k} \underline\alpha^M_{k} - c_{i, k}, 0\}
\]
where the first equation is by that $(M, \alpha^M)$ forms a feasible matching profile. 
This completes the proof.
\end{proof}

By utilizing \cref{minimum_payment} and \cref{minimum_payment_extend}, we construct an algorithm to find the minimum matching $\mathcal{B}_e$, as depicted in Algorithm~\ref{alg:closure}.

\begin{algorithm}[H]
\caption{Computing the minimum matching $\mathcal{B}_e$}
\label{alg:closure}
\KwIn{An assignment $e=(i,j)$, an graph $\mathcal{G}=(\mathcal{A}\cup\mathcal{T}, \mathcal{E})$.}
\KwOut{The closure $\mathcal{B}_e$, or $\bot$ if infeasible.}

$M\gets\{e\}$\;

\While{\textnormal{true}}{ \label{while_loop}
    Solve the linear program
    \begin{equation}\label{LP_computingBe}
    \begin{aligned}
                \min_{1 \ge \alpha \ge 0} \quad &
        \sum_{t\in\mathcal{T}(M)}\alpha_t \\
        \text{subject to} \quad &  \eqref{eq:owner_constraints}
    \end{aligned}
    \end{equation}

    \If{the \eqref{LP_computingBe} is infeasible}{
        \Return $\bot$\;
    }

    Let $\alpha^M$ be an optimal solution of \eqref{LP_computingBe}, and $F\gets\emptyset$\;

    \ForEach{$a\in\mathcal{A}\setminus\mathcal{A}(M)$}{
        \If{$\exists\,t\in\mathcal{T}(M)$ such that 
             $U_{a,t}(\alpha_t^M)>0$}{ \label{strictenvy}
            $F\gets F\cup\{a\}$\;
        }
    }

    \If{$F=\emptyset$}{ \label{line_f_empty}
        \Return $\mathcal{B}_e=M$\;
    }

    \ForEach{$a\in F$}{
        Choose $t\in\mathcal{T}(M)$ such that
        $U_{a,t}(\alpha_t^M)>0$\;

        $t'\gets\operatorname{opp}((a,t))$\;

        \If{$t'$ does not exist or
             $t'\in\mathcal{T}(M)$}{\label{form_new_matching_1}
            \Return $\bot$\;
        }

        $M\gets M\cup\{(a,t')\}$\; \label{form_new_matching_2}
    }
}
\end{algorithm}

The next lemma is ready to show that Algorithm~\ref{alg:closure} returns a minimum matching $\mathcal{B}_e$ if it exists.

\begin{lemma}\label{mathcingincontatins}
Algorithm~\ref{alg:closure} either correctly determines that no feasible matching contains $e$, or returns a minimum matching profile $(\mathcal{B}_e,\alpha^{\mathcal{B}_e})$. The returned matching is unique. 
\end{lemma}

\begin{proof}
Note that the algorithm either returns $\bot$ or returns $\mathcal{B}_e$. When $\mathcal{B}_e$ is returned, it must be the case that $F=\emptyset$ in Line~\ref{line_f_empty}, which implies that there is no envy among agents. The \eqref{eff_constraint} are explicitly imposed in Program \eqref{LP_computingBe}. Hence, the returned $\mathcal{B}_e$ coupled with the obtained contract $\alpha^{\mathcal{B}_e}$  form a feasible matching profile.

Next, we prove that the found $(\mathcal{B}_e, \alpha^{\mathcal{B}_e})$ is the minimum matching profile. For any feasible matching $M$ that contains $e=(i, j)$, we show $\mathcal{B}_e \subseteq M$. Suppose that there exists one edge $l = (p, q) \in \mathcal{B}_e$ but $l \notin M$. Let $U_k$ be the set of edges added to $\mathcal{B}_e$ at the first $k$ iterations (the first $k$ while loops at Line~\ref{while_loop}). Let $l$ be added to $\mathcal{B}_e$ at some iteration $k$ (the $k$ while loop at Line~\ref{while_loop}). Hence, $U_k\subseteq \mathcal{B}_e $ but $U_k\nsubseteq M$. By Program \eqref{LP_computingBe} and \cref{minimum_payment}, we know that at iteration $k-1$, the returned contract $\alpha^{\mathcal{B}_e,k-1}$ from solving Program \eqref{LP_computingBe} has the minimum possible payments for all tasks in $j \in  \mathcal{T}(U_{k-1})$. Therefore, it must hold that $\alpha^M_j \ge \alpha^{\mathcal{B}_e,k-1}_j$ for all the tasks $j \in  \mathcal{T}(U_{k-1})$. This must cause envy from agent $p \in l$ in matching $M$, implying that edge $l$ must be included in $M$. Hence, $\mathcal{B}_e\subseteq M$ must hold, and $\mathcal{B}_e$ is a minimum matching. The uniqueness comes from the fact that the sum of minimum payments must be unique, by \cref{minimum_payment}. This completes the proof.
\end{proof}

\begin{figure}[t]
    \centering
    \includegraphics[width=0.8\textwidth]{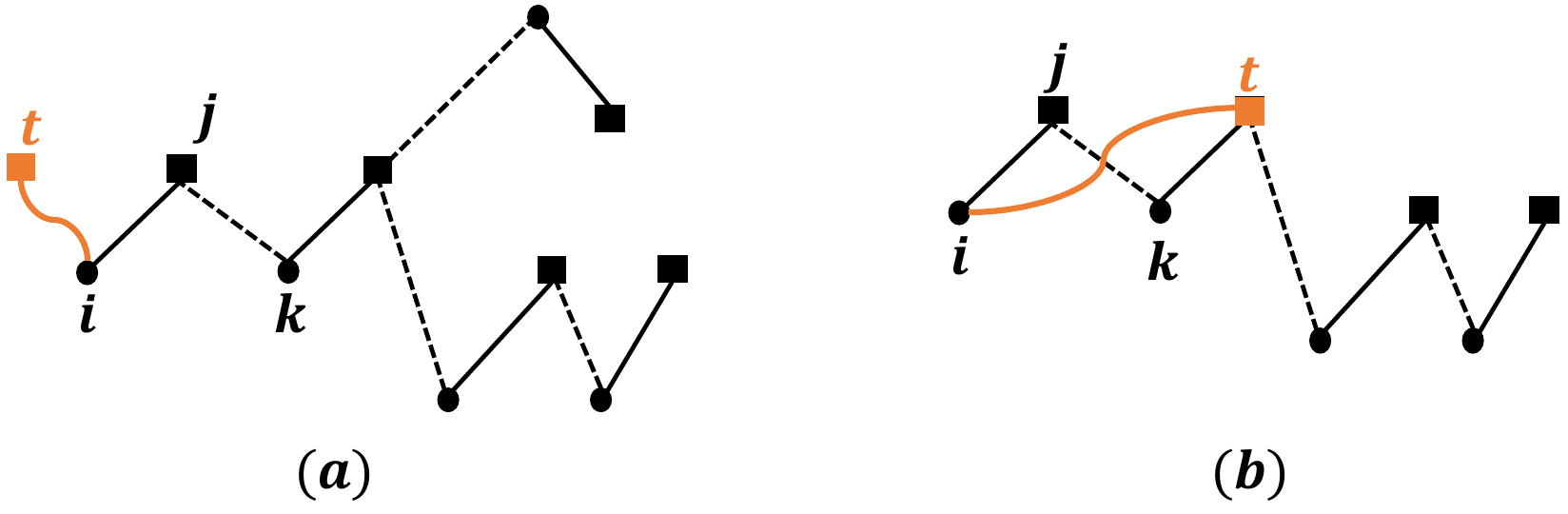}
    \caption{Illustration of the structure of $\mathcal{B}_{(i,j)}$.
Circles represent agents, and squares represent tasks.
Black solid edges indicate task assignments, while black dashed
edges connect agents to their parent tasks.
The orange edge indicates that agent $i$ is interested in task $t$.
Here, $(i,j)$ is the base assignment, and agent $i$ is interested
in both tasks $j$ and $t$.
(a) Task $t$ is unassigned, and the underlying graph of
$\mathcal{B}_{(i,j)}$ forms a tree.
(b) Task $t$ is assigned to agent $k$. Adding the orange edge
between $i$ and $t$ creates a unique cycle containing
the assignments $(i,j)$ and $(k,t)$.}
    \label{fig_my_image}
\end{figure}

In the rest of the section, we call $e$ a base assignment. If $(i,j)\ne e$ was inserted because $i$ envied the assigned task $t$, call $t$ its \emph{parent task}.
The next lemma shows that the minimum matching $\mathcal{B}_e$ is a tree, where the node is the matching edge in $\mathcal{B}_e$, and an edge is added due to agent $i$ envying its parent task $t$, as shown in Figure~\ref{fig_my_image}. Note that there may exist cases where a newly included agent is interested in the other interested task of agent $i$, e.g., task $t$ in Figure~\ref{fig_my_image} (b).  In this case, $\mathcal{B}_e$ is still a tree, as agent $k$'s parent task is task $j$ while the agent $i$ in the base assignment $e$ does not have parent task according to our definition.
\begin{lemma}\label{tree_structione}
    By considering each edge as a node, the minimum matching $\mathcal{B}_e$ induced by edge $e=(i, j)$ is a directed tree rooted at edge $e$
\end{lemma}
\begin{proof}
    Let each matching edge in $\mathcal{B}_e$ be one node. For any two matching edges (nodes) $(p, q)$ and $(k, t)$, draw an arc to connect  $(p, q)$ to $(k, t)$ because either task $q$ is the parent task of agent $k$ or task $t$ is the parent task of agent $p$.
    
    Initially, there is one node $e=(i, j)$ in $\mathcal{B}_e$ . In each round of the while loops in Algorithm~\ref{alg:closure}, if the set $F$ is not empty, it means that all the agents in $k \in F$ must envy some tasks that are already assigned. To eliminate the envy, the algorithm must assign another interested task (denoted as $p \in \mathcal{T}$) to agent $k$. Hence, a node $(k, p)$ is added, and a new arc connects some existing node to node $(k, p)$. Since every new arc is used to connect to a newly constructed node, the structure is a tree.
\end{proof}

A straightforward corollary is that if the other interested task $t$ of agent $i$ in the base assignment $e = (i, j)$ is assigned to some other agent, e.g., agent $k$, in $\mathcal{B}_e$ (see Figure~\ref{fig_my_image} (b)), by connecting node $(i, j)$ and node $(k, t)$, it will form a cycle. 
\begin{corollary}\label{cycle_formation}
    If agent $i$ in the base assignment $e = (i, j)$ has two interested tasks, and both two tasks are assigned in $\mathcal{B}_e$, then connecting these two assignments in $\mathcal{B}_e$ will form a cycle. Moreover, this cycle is unique. 
\end{corollary}

\paragraph{When matchings are compatible.} Given any two matchings $M_1$ and $M_2$, they are either compatible or conflicting. 

\begin{definition}
    We say that two matchings $M_1$ and $M_2$ are conflicting if either there exists one agent $a\in M_1$ and $a\in M_2$ such that it is assigned two different tasks in the two respective matchings (called {\it agent-conflict}), or there exists one task $t\in M_1$ and $t\in M_2$ that is assigned to two different agents in the two respective matchings (called {\it task-conflict}). If two matchings are {\bf not} conflicting, they are compatible.
\end{definition}

We first show the properties for compatible matchings.

\begin{lemma}\label{compatible_sets_laminor}
Consider any two compatible minimum matchings $\mathcal{B}_e$ and $\mathcal{B}_f$ such that $\mathcal{B}_e \cap \mathcal{B}_f \neq \emptyset$.  Let the assignment $h$ be such that $h\in \mathcal{B}_e$ and $h\in \mathcal{B}_f$. We have either one of the following holds:
\[
\mathcal{B}_e \subseteq \mathcal{B}_f\text{ or } \mathcal{B}_f \subseteq \mathcal{B}_e. 
\]
\end{lemma}

\begin{proof}
    Note that if $h=e$ or $h=f$, we know the conclusion holds by \cref{mathcingincontatins}. Next, we consider the case $h\neq e$ and $h \neq f$.

    Let $h = (i,j)$. Since $h$ is not the base assignment, by the iteration of Algorithm~\ref{alg:closure} for finding $\mathcal{B}_e$ and $\mathcal{B}_f$, the task $j$ is assigned to agent $i$ must be due to the fact that its parent task $k$ was assigned, which causes agent $i$'s envy. 
    Since $\mathcal{B}_e$ and $\mathcal{B}_f$ are compatible, task $k$ must be assigned to the same agents (denoted as $p$) in both $\mathcal{B}_e$ and $\mathcal{B}_f$. Then, we have assignment $(p, t) \in \mathcal{B}_e$ and $(p, t) \in \mathcal{B}_f$.
    We repeat the above arguments on $(p, t)$, and finally reach either the base assignment $e$ or $f$, which is included in both $\mathcal{B}_e$ and $\mathcal{B}_f$. Again, by applying \cref{mathcingincontatins}, the conclusion holds.
\end{proof}

Next, we show that for any two compatible minimum matchings, their union also forms a matching, and the contracts to support this matching can be computed by taking the maximum contracts for each task in both contracts.
\begin{lemma}\label{lemma_maximum_contract}
    Consider any set of compatible minimum matchings $\mathcal{B}_e, \forall e \in E$ that are induced by the set of edge $E$. $(M, \alpha^M)$ forms a feasible matching profile, where $M= \cup_{e \in E} \mathcal{B}_e$ and $\alpha^M_j = \max_{e\in E}  \alpha_j^{\mathcal{B}_e}$ for each task $j\in \mathcal{T}$.
\end{lemma}
\begin{proof}
Consider any two minimum matchings $\mathcal{B}_e, \mathcal{B}_f$. 
    If $\mathcal{B}_e \cap \mathcal{B}_f \neq \emptyset$, then \cref{compatible_sets_laminor} implies the nclusion relation. Without loss of generality, assume  $\mathcal{B}_e \subseteq \mathcal{B}_f$. Hence, the minimum payment property of computing minimum matching (\cref{minimum_payment}) immediately implies the conclusion for this case, i.e., $\mathcal{B}_f = \mathcal{B}_e \cup \mathcal{B}_f$ and $\alpha^{\mathcal{B}_f}_j = \max \{\alpha^{\mathcal{B}_e}_j, \alpha^{\mathcal{B}_f}_j\}$. 

    Indeed, for the above case, we can safely remove $\mathcal{B}_e$. By applying the above arguments repeatedly, only a set of non-overlapping matchings will be left. Hence, we will be left $E'\subseteq E$ such that for any two edges $e, f \in E'$, it holds that $\mathcal{B}_{e} \cap \mathcal{B}_{f} = \emptyset$.

    Consider two minimum matchings $\mathcal{B}_e, \mathcal{B}_f$ such that  $\mathcal{B}_e \cap \mathcal{B}_f = \emptyset$.
    We only need to verify that for any two matching edges $(i, j) \in \mathcal{B}_e$ and $(p, q) \in \mathcal{B}_f$, there is no envy between agents $i$ and $p$. For agent $i$, we have that 
    \[
    U_{i,j}(\alpha^M_j) =  U_{i,j}(\alpha^{\mathcal{B}_e}_j) \ge 0 \ge  U_{i, q}(\alpha^{\mathcal{B}_f}_q) = U_{i, q}(\alpha^M_q) 
    \]
    where the first inequality is by \eqref{eff_constraint} and the second inequality is by that agent $i$ does not envy agent $p$ in matching $\mathcal{B}_f$. Similarly, we have the EF constraints hold for agent $p$,  
    \[
    U_{p,q}(\alpha^M_q) =  U_{p,q}(\alpha^{\mathcal{B}_f}_q) \ge 0 \ge U_{p, j}(\alpha^{\mathcal{B}_e}_j) = U_{p, j}(\alpha^M_j). 
    \]
    
The above arguments can be easily extended to cases of multiple non-overlapping matchings. This completes the proof.
\end{proof}

Now, we are ready to show that computing the welfare-maximizing solution to the problem \eqref{program-welfare} can be reduced to finding a maximum weight independent set in a conflict graph $\mathcal{C}$.

\begin{definition}
    The conflict graph $\mathcal{C}$ is defined as follows: For each edge $e=(i,j) \in \mathcal{E}$, if the minimum matching $\mathcal{B}_e$ exists, then construct a vertex $v_e$ with weight $w_e = W_{i, j}$. For any two vertices $v_e, v_f \in \mathcal{C}$, if the two minimum matchings $\mathcal{B}_e$ and $\mathcal{B}_f$ are conflicting, then connect these two vertices by edge $(v_e, v_f)$.
\end{definition}

\begin{proposition}\label{reduction_polynotimal}
    Computing the welfare-maximizing solution to the problem \eqref{program-welfare} can be reduced to finding a maximum weight independent set in the graph $\mathcal{C}$.
\end{proposition}
\begin{proof}
    Recall that a feasible solution to the problem \eqref{program-welfare} corresponds to a feasible matching profile $(M, \alpha^M)$. Consider a matching edge $e \in M$. By the definition of minimum matching, we know that the minimum matching $\mathcal{B}_e$ must be included in $M$. Hence, $\cup_{e\in M} \mathcal{B}_e \subseteq M$. Moreover, we also have $M \subseteq \cup_{e\in M} \mathcal{B}_e$, implying that $M = \cup_{e\in M} \mathcal{B}_e$.

    Since $M$ is a feasible matching, this implies that all the minimum matchings $\mathcal{B}_e, e \in M$ are compatible. Therefore, the set of vertices $\{ v_e\}_{e\in M}$ form an independent set in the conflict graph $\mathcal{C}$. By our construction of the conflict graph, the welfare-maximizing solution hence corresponds to the maximum weight independent set.
\end{proof}

\paragraph{When matchings are conflicting.} We first define the opposite assignment of matching edge $e=(i, j)$. 
\begin{definition}
    Let $opp(e)$ be the opposite assignment of matching edge $e=(i, j)$: Suppose an agent $i$ is interested in two tasks $j, k$. If $e=(i, j)$, then $opp(e) = (i, k)$. If agent $i$ is interested in only one task, then $opp(e)$ is not defined.
\end{definition}
For any two conflicting minimum matchings $\mathcal{B}_e$ and $\mathcal{B}_f$, there are $4$ possible conflicting relations: 
\begin{itemize}
    \item (1)  $opp(e) \in \mathcal{B}_f$, but $opp(f) \notin \mathcal{B}_e$;
    \item (2) $opp(f) \in \mathcal{B}_e$, but $opp(e) \notin \mathcal{B}_f$;
    \item (3) $opp(e) \in \mathcal{B}_f$ and  $opp(f) \in \mathcal{B}_e$;
    \item (4) $opp(e) \notin \mathcal{B}_f$ and  $opp(f) \notin \mathcal{B}_e$.
\end{itemize}   
Note that if $opp(f)$ is not defined, then it holds that $opp(f) \notin \mathcal{B}_e$. Similarly, if $opp(e)$ is not defined, then it holds that $opp(e) \notin \mathcal{B}_f$.

\begin{lemma}\label{single_direction_conf}
    Given any two conflicting minimum matchings $\mathcal{B}_e$ and $\mathcal{B}_f$, if $opp(e) \in \mathcal{B}_f$, but $opp(f) \notin \mathcal{B}_e$, then 
    \[
    \mathcal{T}(\mathcal{B}_e) \subsetneq \mathcal{T}(\mathcal{B}_f), \mathcal{A}(\mathcal{B}_e) \subsetneq \mathcal{A}(\mathcal{B}_f)
    \]
\end{lemma}
\begin{proof}
  Let the base assignment be $e=(a_1, t_1)$ and the opposite assignment be $h=opp(e) = (a_1, t_0)$.  To ease the exposition, we first make an assumption: task $t_0$ is not assigned to any agent in $\mathcal{B}_e$.

Since $opp(f) \notin \mathcal{B}_e$, this implies that $h$ cannot be the base assignment of $\mathcal{B}_f$, i.e., $h \neq f$. Since agent $a_1$ is $2$-interest,   task $t_1$ must be assigned before task $t_0$, i.e., $t_1$ is the agent $a_1$'s parent task in $\mathcal{B}_f$. Then, there are two possible cases: 
    \begin{itemize}
        \item (1) task $t_1$ is assigned to some agent $a_2$ but $a_2$ is not assigned any task in $\mathcal{B}_e$; 
    \item (2) task $t_1$ is assigned to some agent $a_2$ and agent $a_2$ is also assigned some task $t_2$ in $\mathcal{B}_e$. 

    \end{itemize}

    Consider case (2) holds.  Then, we know that the assignment $(a_2, t_1)$ cannot be the base assignment $f$; otherwise, we have $opp(f)\in \mathcal{B}_e$ due to $a_2$ is $2$-interest, a contradiction. 
    Hence, task $t_2$ is the parent task of agent $a_2$ in $\mathcal{B}_f$. 
    Since $opp(f)\notin \mathcal{B}_e$, by repeating the above process, we will finally reach some agent $a_k$ in $\mathcal{B}_f$ such that it is assigned some task $t_{k-1}$, but agent $a_k$ is not assigned any task in $\mathcal{B}_e$. Therefore, the problem reduces to case (1). 

Now, we consider case (1). By the above arguments, 
 we assume that it will finally reach a situation where each task in set $\{t_1, t_2, \dots, t_{k-1}\}$ is respectively assigned to each agent $\{a_1, a_2, \dots, a_{k-1}\}$ in $\mathcal{B}_e$ (i.e., $(a_i, t_i) \in \mathcal{B}_e$ for $i \in \{1, 2, \dots, k-1\}$), while task $t_j \in \{t_1, t_2, \dots, t_{k-1}\}$ is assigned to agent $a_{j+1}$ and agent $a_1$ is assigned task $t_0$ in $\mathcal{B}_f$ (i.e., $(a_{i+1}, t_{i}) \in \mathcal{B}_f$ for $i \in \{1, 2, \dots, k-1\}$). 
 Moreover, agent $a_{k}$ is not matched in $\mathcal{B}_e$. Therefore, by EF constraints for agent $a_{k}$ in $\mathcal{B}_e$ and effort constraints for agent $a_{k}$ in $\mathcal{B}_f$, we know that 
\[
\alpha^{\mathcal{B}_e}_{t_{k-1}} p_{a_{k}, t_{k-1}} r_{t_{k-1}} - c_{a_{k}, t_{k-1}} \le 0 \le \alpha^{\mathcal{B}_f}_{t_{k-1}} p_{a_{k}, t_{k-1}} r_{t_{k-1}} - c_{a_{k}, t_{k-1}}
\]
implying $\alpha^{\mathcal{B}_f}_{t_{k-1}} \ge \alpha^{\mathcal{B}_e}_{t_{k-1}}$. Next, consider task $t_{k-2}$. By EF constraints, we have 
\begin{align*}
    \alpha^{\mathcal{B}^f}_{t_{k-2}} p_{a_{k-1}, t_{k-2}} r_{t_{k-2}} - c_{a_{k-1}, t_{k-2}} &\ge \alpha^{\mathcal{B}^f}_{t_{k-1}} p_{a_{k-1}, t_{k-1}} r_{t_{k-1}} - c_{a_{k-1}, t_{k-1}} \\
    \alpha^{\mathcal{B}^e}_{t_{k-2}} p_{a_{k-1}, t_{k-2}} r_{t_{k-2}} - c_{a_{k-1}, t_{k-2}} &\le \alpha^{\mathcal{B}^e}_{t_{k-1}} p_{a_{k-1}, t_{k-1}} r_{t_{k-1}} - c_{a_{k-1}, t_{k-1}} 
\end{align*}
together with $\alpha^{\mathcal{B}_f}_{t_{k-1}} \ge \alpha^{\mathcal{B}_e}_{t_{k-1}}$, implying that $\alpha^{\mathcal{B}^f}_{t_{k-2}} \ge \alpha^{\mathcal{B}^e}_{t_{k-2}}$. By repeatedly applying similar arguments, we have that $\alpha^{\mathcal{B}_f}_{t_{j}} \ge \alpha^{\mathcal{B}_e}_{t_{j}}$ holds for all the tasks in $t_j \in \{t_1, t_2, \dots, t_{k-1}\}$.


Recall that the minimum matching $\mathcal{B}_e$ is indeed a tree (\cref{tree_structione}). Since the tasks in $\{t_1, t_2, \dots, t_{k-1}\}$ are sequentially assigned starting from the base assignment $e$ in $\mathcal{B}_e$, the assignments $(a_1, t_1), (a_2, t_2), \dots, (a_{k-2}, t_{k-2}), (a_{k-1}, t_{k-1})$ forms a path rooted at $e$. 
A branch in $\mathcal{B}_e$ connecting to the above path is added due to that the agent in the root assignment of that branch envies some task in $\{t_1, t_2, \dots, t_{k-1}\}$. 
 Since $\alpha^{\mathcal{B}_f}_{t_{j}} \ge \alpha^{\mathcal{B}_e}_{t_{j}}$ hold for all the tasks in $t_j \in \{t_1, t_2, \dots, t_{k-1}\}$, the agent in the root assignment of that branch continues to envy. Hence, those branches are included in $\mathcal{B}_f$ as well. Hence, we have 
\[
    \mathcal{T}(\mathcal{B}_e) \subseteq \mathcal{T}(\mathcal{B}_f), \mathcal{A}(\mathcal{B}_e) \subseteq  \mathcal{A}(\mathcal{B}_f)
    \]
 Moreover, since task $t_0\notin \mathcal{B}_e$ and agent $a_{k} \notin \mathcal{B}_e$, we know that 
\[
    \mathcal{T}(\mathcal{B}_e) \subsetneq \mathcal{T}(\mathcal{B}_f), \mathcal{A}(\mathcal{B}_e) \subsetneq  \mathcal{A}(\mathcal{B}_f).
    \]

    Finally, to complete the proof, we show that the assumption that task $t_0$ is not assigned to any agent in $\mathcal{B}_e$ holds by the conditions of the lemma. 
Assume by contradiction that task $t_0$ is assigned to some agent in $\mathcal{B}_e$. Divide the analysis into two cases: (1) $t_0$ is included in an off-path branch; and (2) $t_0$ is included in the above path. 

In case (1), assume task $t_0$ is assigned to agent $a_{t_0}  \neq a_1$ in $\mathcal{B}_e$. By the above argument, we know that this off-path branch will be included in $\mathcal{B}_f$. However, task $t_0$ is assigned to agent $a_1$ in $\mathcal{B}_f$, leading to a contradiction to the claim that $\mathcal{B}_f$ is a feasible matching. 

In case (2), suppose that now $(a_1, t_1), (a_2, t_2), \dots, (a_{d}, t_{d})$ is part of the path in $\mathcal{B}_e$, where $t_d = t_0$. By the above arguments, we know that agent $a_d$ is assigned task $t_{d-1}$ in $\mathcal{B}_f$. If $(a_d, t_{d-1})$ is the base assignment $f$, then $opp(f) \in \mathcal{B}_e$, contradicting the conditions of the lemma; Otherwise, since $(a_d, t_0) \in \mathcal{B}_e$, we know that agent $a_d$ is interested in tasks $t_0$ and $t_{d-1}$. Note that $(a_1, t_0) \in \mathcal{B}_f$. Therefore, $(a_1, t_0), (a_2, t_1), \dots, (a_d, t_{d-1})$ form a cycle in $\mathcal{B}_f$. By \cref{cycle_formation}, we know that $f$ must belong to this cycle, implying that $opp(f) \in \mathcal{B}_e$, contradicting the condition of the lemma.
\end{proof}

\begin{lemma}\label{lemmaconflict_task}
    Given any two conflicting minimum matchings $\mathcal{B}_e$ and $\mathcal{B}_f$, if  $opp(e) \notin \mathcal{B}_f$ and $opp(f) \notin \mathcal{B}_e$, then 
    \[
    \mathcal{T}(\mathcal{B}_e) = \mathcal{T}(\mathcal{B}_f), \mathcal{A}(\mathcal{B}_e)\setminus\{a_e\} = \mathcal{A}(\mathcal{B}_f)\setminus\{a_f\}
    \]
\end{lemma}

\begin{proof}

Let the base assignments be $e=(a_e, t_e)$ and $f=(a_f, t_f)$.

    Assume that the two matchings are of agent-conflict. 
    If $\mathcal{B}_e$ and $\mathcal{B}_f$ have conflicts at the agent $a_e$ or agent $a_f$, it implies that either $opp(e) \in \mathcal{B}_f$ or $opp(f) \in \mathcal{B}_e$ holds, contradicting the assumption of the lemma. 
    Hence, suppose matchings $\mathcal{B}_e$ and $\mathcal{B}_f$ conflict at agent agent $a$ where $a\neq a_e$ and $a\neq a_f$. Let the assignments in $\mathcal{B}_e$ and $\mathcal{B}_f$ be $(a, t_a^e)$ and $(a, t^f_a)$, respectively.
    Since $(a, t_a^e)$ is not the base assignment, this implies that task $t^f_a$ is the parent task of agent $a$ and is assigned immediately to some agent other than $a$ in $\mathcal{B}_e$. Then, the conflict type of $\mathcal{B}_e$ and $\mathcal{B}_f$ becomes task-conflict; the two matchings have conflicts in task $t^f_a$.

    In the following, we consider the matchings of task-conflict. Assume that both matchings have a conflict at task $t$ where $(a, t) \in \mathcal{B}_e$.  Then, for the agent $b$ assigned task $t$ in $\mathcal{B}_f$, there are two possible cases: 
    \begin{itemize}
        \item (1) the agent is also assigned a task in $\mathcal{B}_e$;
        \item (2) the agent is not assigned any task in $\mathcal{B}_e$.
    \end{itemize}

    Consider case (2). Since agent $b$ is unmatched in $\mathcal{B}_e$, 
it implies that the contract $\alpha^f_t$ for task $t$ in $\mathcal{B}_f$ is weakly higher,  $\alpha^f_t \ge \alpha^e_t$, by effort constraints and EF constraints.

If $\alpha^f_t > \alpha^e_t$ strictly holds and task $t$ is a parent task of agent $a$ in $\mathcal{B}_f$, the agent $a$ at least gains $\alpha^f_t p_{a, t} r_t - c_{a, t} > \alpha^e_t p_{a, t} r_t - c_{a, t}$ in $\mathcal{B}_f$ . 
Since $(a, t)$ is not the base assignment of $\mathcal{B}_e$, we know that 
agent $a$ is interested in task $t_1 \neq t$. 
Hence, in $\mathcal{B}_f$, we know that
 $\alpha^f_{t_1} p_{a, {t_1}} r_{t_1} - c_{a, {t_1}} \ge \alpha^f_t p_{a, t} r_t - c_{a, t} > \alpha^e_t p_{a, t} r_t - c_{a, t} \ge \alpha^e_{t_1} p_{a, {t_1}} r_{t_1} - c_{a, {t_1}}$, implying $\alpha^f_{t_1} > \alpha^e_{t_1}$. Repeating this argument along the path back to the base $(a_e, t_e)$ in $\mathcal{B}_e$, 
 we have $\alpha^f_{t_e} > \alpha^e_{t_e}$. This implies that to satisfy the EF constraints, agent $a_e$ must be assigned an interested task other than $t_e$ in $\mathcal{B}_f$. This implies that $opp(e)\in \mathcal{B}_f$, contradicting the assumption of the lemma. Therefore, this case cannot happen.
    
    Therefore, if case (2) holds, it must be that $\alpha^f_t = \alpha^e_t$. This indicates that $\alpha^f_t p_{b, t} r_t - c_{b, t} = 0$, which further implies that $(b, t)$ is the base assignment of $\mathcal{B}_f$ by Line~\ref{strictenvy} in Algorithm~\ref{alg:closure} that each non-base assignment gives strictly positive utility due to strict envy. Let the path from the base assignment $(a_e, t_e)$ to $(a, t)$ in $\mathcal{B}_e$ be $(a_e, t_e), (a_k, t_k), (a_{k-1}, t_{k-1}), \dots, (a_1, t_1), (a, t)$. Note that agent $a$ envies task $t$ in $\mathcal{B}_f$, to satisfy EF constraints, agent $a$ must be assigned task $t_1$. Then, $\alpha^{f}_{t_1} \ge \alpha^{e}_{t_1}$ since 
    \[
    \alpha^f_{t_1} p_{a, {t_1}} r_{t_1} - c_{a, {t_1}} \ge \alpha^f_t p_{a, t} r_t - c_{a, t} = \alpha^e_t p_{a, t} r_t - c_{a, t} \ge \alpha^e_{t_1} p_{a, {t_1}} r_{t_1} - c_{a, {t_1}}.
    \]
    Suppose $\alpha^{f}_{t_1} > \alpha^{e}_{t_1}$ strictly holds. We know that agent $a_1$ must be assigned task $t_2$ in  $\mathcal{B}_f$ to satisfy EF constraints. Then, we have $\alpha^{f}_{t_2} > \alpha^{e}_{t_2}$. Repeating this argument along the path, we have $\alpha^f_{t_e} > \alpha^e_{t_e}$, leading to $opp(e)\in \mathcal{B}_f$ as in the above case where $\alpha^f_t > \alpha^e_t$ strictly holds. Hence, $\alpha^{f}_{t_1} = \alpha^{e}_{t_1}$. By induction, we finally have $\alpha^{f}_{t_j} = \alpha^{e}_{t_j}$ for all the tasks on the path $t_j = t, t_1, t_2, \dots,t_k, t_e$. Moreover, all the off-path branches of $\mathcal{B}_e$ must also be included in $\mathcal{B}_f$, since the contracts for the tasks on the path do not change, which indicates that the agents in the root assignments of off-path branches still envy tasks on the path, inducing the assignments in the branches. Now, we consider agent $a_e$. We show that $U_{a_e, t_e}(\alpha^f_{t_e}) = 0$. Then, we have that the conclusion $\mathcal{T}(\mathcal{B}_e) = \mathcal{T}(\mathcal{B}_f), \mathcal{A}(\mathcal{B}_e)\setminus\{a_e\} = \mathcal{A}(\mathcal{B}_f)\setminus\{a_f\}$ holds. If $U_{a_e, t_e}(\alpha^f_{t_e}) > 0$, agent $a_e$ must be assigned an interested task other than $t_e$ in $\mathcal{B}_f$. This implies that $opp(e)\in \mathcal{B}_f$, contradicting the assumption of the lemma.

    Now, consider case (1). Suppose that agent $b$ is assigned task $t_{-1}$ in $\mathcal{B}_e$. If $(b, t)$ is the base assignment of $\mathcal{B}_f$, then we have $opp(f) \in \mathcal{B}_e$, contradicting the assumption. Hence, $(b, t)$ cannot be the base assignment, and task $t_{-1}$ is the parent task of agent $b$ and is assigned immediately before task $t$. Suppose that task $t_{-1}$ is assigned to some agent $b_{-1}$ in $\mathcal{B}_f$. If $b_{-1}$ is not assigned any task in $\mathcal{B}_e$, we reach the conditions of case (2) and can apply the same arguments. If $b_{-1}$ is assigned a task in $\mathcal{B}_e$, we again reach the condition of case (1). Hence, by repeatedly applying the argument, we will finally reach the last task $t_{-k}$ of the path in $\mathcal{B}_e$, which is assigned to agent $b_{-(k-1)}$, and agent $b_{-(k-1)}$ is assigned task $t_{-(k-1)}$ in $\mathcal{B}_f$. We know that $(b_{-(k-1)}, t_{-(k-1)})$ cannot be the base assignment of $\mathcal{B}_f$; otherwise, $opp(f) \in \mathcal{B}_e$, contradicting the assumption of the lemma. Hence, task $t_{-k}$ must be assigned to some agent $b_{-k}$ in $\mathcal{B}_f$. Moreover, since task $t_{-k}$ is the last task on the path, which is not matched in $\mathcal{B}_e$, we again reach the condition of case (2). By the above arguments for case (2), we have the conclusion holds. This completes the proof.
\end{proof}

\paragraph{Conflict graph $\mathcal{C}$ is a perfect graph.} In this part, we prove an important property of the constructed conflict graph $\mathcal{C}$: $\mathcal{C}$ is a perfect graph~\citep{golumbic2004algorithmic}.

\begin{definition}
A graph $G$ is a chordless cycle on $k$ distinct vertices $v_0, v_1, \dots, v_{k-1}$ if there is an edge from $v_i$ to $v_{i+1}$ mod $k$ (for all $i = 0, 1, \dots, k-1$) and no other edge between any two of these vertices.  A hole is a chordless cycle on five or more
vertices; an antihole is the complement of a hole. 
\end{definition}

\begin{definition}
    Let $H_n$ be a hole composed of $n$ vertices, and co-$H_n$ be the antihole of $H_n$. A graph is a Berge graph if it  does not contain $H_n$ and co-$H_n$ for odd numbers $n=5, 7, 9,\dots$. 
\end{definition}

In this part, we mainly prove the following key theorem. 

\begin{theorem}\label{perfect_graph}
    The conflict graph $\mathcal{C}$ is a perfect graph.
\end{theorem}

By the seminal Strong Perfect Graph Theorem~\citep{chudnovsky2006strong}, a graph is a perfect graph if and only if it is a Berge graph.
Therefore, we only need to show that the conflict graph $\mathcal{C}$ is a Berge graph. Before presenting the proof, we show several useful lemmas. Recall that each node in $\mathcal{C}$ corresponds to a base assignment that induces an minimum matching. In the following, we use vertex to refer a vertex in a graph and a base assignment interchangeably. 

\begin{lemma}\label{non_adj_no_overl}
    For any two non-adjacent vertices $i, j$ in odd hole $H_n$ or odd antihole co-$H_n$, $n=5, 7, 9...$, their sets of agents and tasks in the minimum matchings respectively induced by base assignments $i$ and $j$  do not overlap. In other words,
    \[
    \mathcal{T}(\mathcal{B}_i) \cap \mathcal{T}(\mathcal{B}_j) = \emptyset, \mathcal{A}(\mathcal{B}_i) \cap \mathcal{A}(\mathcal{B}_j) = \emptyset.
    \]
\end{lemma}
\begin{proof}
    We prove the lemma by contradiction. 
Assume by contradiction that $\mathcal{A}(\mathcal{B}_i) \cap \mathcal{A}(\mathcal{B}_j) \neq \emptyset $. Let agent $a \in \mathcal{A}(\mathcal{B}_i) \cap \mathcal{A}(\mathcal{B}_j)$. Since vertices $i, j$ are non-adjacent, it means that the matchings $\mathcal{B}_i$ and $\mathcal{B}_j$ are compatible.
This further implies that 
the assigned tasks to agent $a$ must be the same in $\mathcal{B}_i$ and $\mathcal{B}_j$, denoted as $t$. 
By \cref{compatible_sets_laminor}, we know that either $\mathcal{B}_i\subseteq \mathcal{B}_j$ or $\mathcal{B}_j \subseteq \mathcal{B}_i$ holds. Without loss of generality, we assume $\mathcal{B}_i\subseteq \mathcal{B}_j$ hold.

Consider an odd hole $H_n$ with $n=5, 7, \dots$. For any two non-adjacent vertices $i$ and $j$, one can always find a vertex $k \in H_n$ such that edge $(k, i) \in H_n$ but  $(j,k) \notin H_n$. Therefore, $\mathcal{B}_k$ and $\mathcal{B}_i$ have conflict. Since $\mathcal{B}_i\subseteq \mathcal{B}_j$, we have $\mathcal{B}_k$ and $\mathcal{B}_j$ have conflict. However, since $(j,k) \notin H_n$, we have $\mathcal{B}_k$ and $\mathcal{B}_j$ are compatible, reaching a contradiction. Hence, $\mathcal{A}(i) \cap \mathcal{A}(j) = \emptyset$. 

Consider an odd antihole co-$H_n$ with $n=5, 7, \dots$. We label the vertices according to the hole $H_n$ as $v_0, v_1, \dots, v_{n-1}$ where $v_0$ is adjacent to $v_{n-1}$ in hole $H_n$. By definition of antihole, any two non-adjacent vertices $v_i$ and $v_{j}$ in co-$H_n$ satisfy $j= i+1 \text{~mod~} n$. Observe that we can find the vertex $k = j+1 \text{~mod~} n$ such that edge $(k, i) \in H_n$ but  $(j,k) \notin H_n$. Then, by a similar argument as for the case of odd hold $H_n$, we can show $\mathcal{A}(\mathcal{B}_i) \cap \mathcal{A}(\mathcal{B}_j) = \emptyset$. 

Similarly, we can also show $\mathcal{T}(\mathcal{B}_i) \cap \mathcal{T}(\mathcal{B}_j) = \emptyset$ by contradiction. This completes the proof.
\end{proof}

Next, we show that the single-direction relation cannot exist in an odd hole or an odd antihole.
\begin{lemma}\label{single_direction_not_exist}
    For any two adjacent vertices $v_i,v_j$ in an odd hole $H_n$ or an odd antihole co-$H_n$, $n=5, 7, 9...$, the single-direction relations cannot exist, i.e., both (1)  $opp(v_i) \in \mathcal{B}_{v_j}$, but $opp(v_j) \notin \mathcal{B}_{v_i}$, and (2) $opp(v_j) \in \mathcal{B}_{v_i}$, but $opp(v_i) \notin \mathcal{B}_{v_j}$ cannot hold.
\end{lemma}
\begin{proof}
    Without loss of generality, we only need to prove that it cannot hold that $opp(v_i) \in \mathcal{B}_{v_j}$, but $opp(v_j) \notin \mathcal{B}_{v_i}$. We label the vertices according to the hole $H_n$ as $v_0, v_1, \dots, v_{n-1}$ where $v_0$ is adjacent to $v_{n-1}$ in hole $H_n$. The lemma is proved by contradiction.

    Consider an odd hole $H_n$. Choose any two adjacent vertices $v_i$ and $v_{i+1}$. If $i+1 = n$, then choose two vertices $v_{n-1}$ and $v_0$.
    Assume that $opp(v_i) \in \mathcal{B}_{v_{i+1}}$ but $opp(v_{i+1}) \notin \mathcal{B}_{v_{i}}$ holds.  
    \cref{single_direction_conf} thus implies that $\mathcal{T}(\mathcal{B}_{v_i}) \subsetneq \mathcal{T}(\mathcal{B}_{v_{i+1}}), \mathcal{A}(\mathcal{B}_{v_i}) \subsetneq \mathcal{A}(\mathcal{B}_{v_{i+1}})$. 
    Consider vertex $v_{i-1}$ and hence, $(v_i,v_{i-1}) \in H_n$ and $(v_{i-1}, v_{i+1})\notin H_n$. 
    Since vertex $v_{i-1}$ and $v_{i}$ have conflicts, it implies that $\mathcal{B}_{v_{i-1}}$ and $\mathcal{B}_{v_{i}}$ have either agent conflicts or task conflicts, i.e., either $\mathcal{T}(\mathcal{B}_{v_{i}})\cap \mathcal{T}(\mathcal{B}_{v_{i-1}}) \neq \emptyset$ or $\mathcal{A}(\mathcal{B}_{v_{i}})\cap \mathcal{A}(\mathcal{B}_{v_{i-1}}) \neq \emptyset$ holds. 
    Therefore, we have that either $\mathcal{T}(\mathcal{B}_{v_{i+1}})\cap \mathcal{T}(\mathcal{B}_{v_{i-1}}) \neq \emptyset$ or $\mathcal{A}(\mathcal{B}_{v_{i+1}})\cap \mathcal{A}(\mathcal{B}_{v_{i-1}}) \neq \emptyset$ holds.
    However, since edge $(v_{i-1}, v_{i+1})\notin H_n$, \cref{non_adj_no_overl} implies that both $\mathcal{T}(\mathcal{B}_{v_{i+1}})\cap \mathcal{T}(\mathcal{B}_{v_{i-1}}) = \emptyset$ and $\mathcal{A}(\mathcal{B}_{v_{i+1}})\cap \mathcal{A}(\mathcal{B}_{v_{i-1}}) = \emptyset$ hold. We reach a contradiction. Hence, the conclusion holds for odd hole $H_n$.

    Next, we consider an odd antihole co-$H_n$. Consider two adjacent  vertices $v_i,v_j$ in co-$H_n$. Assume that $opp(v_i) \in \mathcal{B}_{v_j}$ but $opp(v_j) \notin \mathcal{B}_{v_i}$ holds. Since $n\ge 5$, there must exist a vertex $v_k$ such that edge $(v_k, v_j)\notin H_n$ and $(v_i, v_k)\in H_n$. Thus, by a similar argument as for the case of odd hole $H_n$, we prove that the conclusion also holds for co-$H_n$.
\end{proof}

Now, we are ready to show the proof of \cref{perfect_graph}.

\begin{proofof}{\cref{perfect_graph}}
    To prove that $\mathcal{C}$ is a Berge graph, we only need to show that graph $\mathcal{C}$ does not contain odd holes $H_n$ or odd antiholes co-$H_n$ for $n=5, 7, 9,\dots$. By \cref{single_direction_not_exist}, we know that given any two conflicting minimum matchings $\mathcal{B}_e$ and $\mathcal{B}_f$ induced by two adjacent vertices $e, f \in \mathcal{C}$, only $2$ possible relations exist in odd holes and odd antiholes, i.e., (1) $opp(e) \in \mathcal{B}_f$ and  $opp(f) \in \mathcal{B}_e$, and (2) $opp(e) \notin \mathcal{B}_f$ and  $opp(f) \notin \mathcal{B}_e$.

    \paragraph{Odd holes $H_n$ do not exist.} Assume by contradiction  that an odd hole $H_n$ exists. 
    
    Consider two consecutive edges $(v_i,v_{i-1}) \in H_n$ and $(v_i,v_{i+1}) \in H_n$, so that $(v_{i-1}, v_{i+1})\notin H_n$. We show that the two consecutive edges cannot have the same type of relation: 
    \begin{itemize}
        \item (i) Let both edges have type (1) relation, i.e., $opp(v_i) \in \mathcal{B}_{v_{i+1}}$ and  $opp(v_{i+1}) \in \mathcal{B}_{v_i}$, and $opp(v_i) \in \mathcal{B}_{v_{i-1}}$ and  $opp(v_{i-1}) \in \mathcal{B}_{v_i}$ hold. 
    Since $(v_{i-1}, v_{i+1})\notin H_n$, \cref{non_adj_no_overl} implies that $\mathcal{T}(\mathcal{B}_{v_{i-1}})\cap \mathcal{T}(\mathcal{B}_{v_{i+1}}) = \emptyset$. 
    However, since $opp(v_i) \in \mathcal{B}_{v_{i+1}}$ and $opp(v_i) \in \mathcal{B}_{v_{i-1}}$, together with that $\mathcal{B}_{v_{i+1}}$ and $\mathcal{B}_{v_{i-1}}$ are compatible, \cref{compatible_sets_laminor} implies that either $\mathcal{B}_{v_{i+1}} \subseteq \mathcal{B}_{v_{i-1}}$ or $\mathcal{B}_{v_{i-1}} \subseteq \mathcal{B}_{v_{i+1}}$ holds, contradicting $\mathcal{T}(\mathcal{B}_{v_{i-1}})\cap \mathcal{T}(\mathcal{B}_{v_{i+1}}) = \emptyset$. 
    \item (ii) Let both edges have type (2) relation, i.e., $opp(v_i) \notin \mathcal{B}_{v_{i+1}}$ and  $opp(v_{i+1}) \notin \mathcal{B}_{v_i}$, and $opp(v_i) \notin \mathcal{B}_{v_{i-1}}$ and  $opp(v_{i-1}) \notin \mathcal{B}_{v_i}$ hold. 
    By \cref{lemmaconflict_task}, we thus know that $\mathcal{T}(\mathcal{B}_{v_i}) = \mathcal{T}(\mathcal{B}_{v_{i+1}}) = \mathcal{T}(\mathcal{B}_{v_{i-1}})$. 
    However, since $(v_{i-1}, v_{i+1})\notin H_n$, \cref{non_adj_no_overl} implies that $ \mathcal{T}(\mathcal{B}_{v_{i+1}}) \cap \mathcal{T}(\mathcal{B}_{v_{i-1}}) = \emptyset$, reaching a contradiction.
    \end{itemize}

For an odd hole $H_n$ with vertices $v_0, v_1, \dots, v_{n-1}$ where $v_0$ is adjacent to $v_{n-1}$, we label the edges with two types of conflict relations alternatively. Finally, we will have edge $(v_0, v_1)$ and edge $(v_0, v_{n-1})$ labeled with the same type of relation, which is impossible by the above arguments. This implies that an odd hole cannot exist.

\paragraph{Odd antiholes co-$H_n$ do not exist.}
    We label the vertices of co-$H_n$ as $v_0, v_1, \dots, v_{n-1}$ according to hole $H_n$ where $v_0$ is adjacent to $v_{n-1}$ in $H_n$. Consider any vertex $v_i$ in co-$H_n$. Select any two vertices $v_j$ and $v_{j+1~\text{mod}~n}$ that are connected to vertex $v_i$. Our first step is to show that edge $(v_i,v_j)$ and edge $(v_i, v_{j+1})$ cannot have the same type of conflict relation. In fact, 
since $(v_j, v_{j+1}) \notin H_n$, by a similar argument as in the case of odd holes, we show that they cannot have type (1) relation or type (2) relation simultaneously.

    Since co-$H_n$ is a complement of $H_n$, for any vertex $v_i \in \text{co-}H_n$, we know that vertex $v_i$ is connected to an even number of vertices in co-$H_n$, i.e., $v_{(i+2)\text{~mod~}n}, v_{(i+3)\text{~mod~}n}, \dots, v_{(i-2)\text{~mod~}n}$. Since vertex $v_i$ is connected to an even number of vertices, it must be that edge $(v_i, v_{(i+2)\text{~mod~}n})$ and edge $(v_i, v_{(i-2)\text{~mod~}n})$ have different types of conflict relation. Now, consider vertex $v_{(i+2)\text{~mod~}n}$, and it is connected to $v_{(i+4)\text{~mod~}n}, v_{(i+5)\text{~mod~}n}, \dots, v_{i \text{~mod~}n}$. Therefore, if we label edge $(v_i, v_{(i+2)\text{~mod~}n})$ as type (1) relation, then the type of conflict for edge $( v_{(i+2)\text{~mod~}n}, v_{(i+4)\text{~mod~}n})$ must be type (2) relation. Consider the following sequence of vertices 
    \[
    v_0, v_1, v_2, \dots, v_{n-1}
    \]
    where $n$ is an odd number. By applying the above key observation, we know that the  edges in sequence $(v_0, v_2), (v_2, v_4), (v_4, v_6), \dots$ are different labeled conflict relation types alternatively. Since $n$ is an odd number, the sequence repeats after $n$ edges, and the last edge of every repetition is $(v_{n-2}, v_0)$. As we label edges with types alternatively, we will have that edges $(v_0, v_2)$ and $(v_{n-2}, v_{0})$ have the same type of relation. This is impossible, as $(v_0, v_2)$ would be the edge immediately after $(v_{n-2}, v_{0})$ in sequence and it should be labeled a type different from that of $(v_{n-2}, v_{0})$. Hence, odd antiholes co-$H_n$ do not exist. 

    Therefore, $\mathcal{C}$ is a perfect graph, completing the proof.
\end{proofof}

\paragraph{Putting all together.} We are now ready to prove \cref{polynotimaitmelaoghtm}. 

\begin{proofof}{\cref{polynotimaitmelaoghtm}} Recall that \cref{reduction_polynotimal} shows that the problem can be reduced to finding a maximum weight independent set in a conflict graph. Moreover, we know that 
the maximum weight independent set in a perfect graph can be computed in polynomial time~\citep{grotschel1981ellipsoid}. After we find a feasible assignment by finding the the maximum weight independent set, we then fix the allocation for each agent and solve the program~\eqref{program-welfare} for the contract $\alpha$, which can be done in polynomial time. Therefore, if we can construct the conflict graph $\mathcal{C}$ in polynomial time, the algorithm is polynomial-time.  Indeed, we can compute the minimum matching for each possible edge $e\in \mathcal{E}$ using Algorithm~\ref{alg:closure}, which is clearly polynomial-time. Since there are at most $nm$ number of edge, there are at most $nm$ vertices in $\mathcal{C}$. It can be seen that checking whether two minimum matchings are compatible needs polynomial time. Hence, we obtain a polynomial-time algorithm for the problem~\eqref{program-welfare}.
\end{proofof}

\section{NP-hardness for $3$-interests Instances }
\label{sec:expanded-welfare-hardness}

In this section, we complement the complexity landscape of problem with unit-demand agents by showing that welfare-maximizing problem becomes $\NP$-hard for $k$-interests instances where $k\ge 3$.

\begin{theorem}
\label{thm:welfare-three-interests}
When agents are unit-demand and have $k\ge 3$ interested tasks, maximizing social welfare is NP-hard. In particular, the following decision problem $\NP$-hard: given a 
 fair-contract instance in which every agent is interested in at most $3$ tasks, and
 a rational threshold $K$, determine whether there exist a matching
 $S$ and contracts $\alpha$ satisfying \eqref{eff_constraint} and \eqref{ef_cons_eq} such that
 $\texttt{Wel}(S, \alpha)\ge K$.
\end{theorem}

\begin{proof}
We reduce from the classical $3$-SAT problem, which is known to be
NP-complete. An instance of $3$-SAT consists of a Boolean formula
$\varphi$ in conjunctive normal form (CNF), where each clause
contains exactly three literals. Formally, let
$\varphi = C_1 \land C_2 \land \cdots \land C_m$ 
be a $3$-CNF formula over Boolean variables
$x_1,\ldots,x_n$, where each clause
$C_t = (\ell_{t,1} \lor \ell_{t,2} \lor \ell_{t,3})$
and each literal $\ell_{t,h}$ is either a variable $x_v$
or its negation $\neg x_v$.
The objective is to determine whether there exists a truth
assignment $\sigma:\{x_1,\ldots,x_n\}\to
\{\mathsf{T},\mathsf{F}\}$ such that every clause
contains at least one true literal.

Given an arbitrary instance $\varphi$ of $3$-SAT,
we construct, in polynomial time, an instance of the
fair-contract welfare maximization problem and a threshold
$K$ such that $\varphi$ is satisfiable if and only if
there exists an individually rational and envy-free
contract with social welfare at least $K$.

 We may assume $n,m\ge1$.
 Put
 \[
 B=n+m+1,\qquad R=8B,\qquad
 \delta=\frac1{8B},\qquad \eta=\frac\delta2,
 \qquad K=\frac n2+\frac{4m}{R}.
 \tag{1}\label{eq:parameters}
 \]
 In particular,
 \[
 0<\eta<\delta<1,\qquad n\delta<1-\eta,
 \qquad \frac12>\frac{2n\delta}{R}.
 \tag{2}\label{eq:parameter-ineq}
 \]
 We construct the following tasks and agents. All agent-task pairs not explicitly
 listed have success probability and cost zero.
 IR, EF, welfare, or revenue.

\paragraph{Normalization and invariance.}
The construction below is obtained by dividing every task reward
and every effort cost in the unnormalized gadget by the same positive
number $R$. All success probabilities and contract shares remain
unchanged. This multiplies every agent's utility and every matching's
social welfare by $1/R$; consequently, it preserves all weak-IR and
EF inequalities and the truth of the decision problem when the
welfare threshold is scaled by $1/R$. We state all quantities below
\emph{on the normalized scale}. In particular, the originally large
root reward becomes $1$, and all the other task rewards are smaller
than $1$.

\paragraph{Construction.}

For each variable $x_v$, introduce two tasks $J_v^{\mathsf T}$ and
 $J_v^{\mathsf F}$, both with reward $1$, and one variable agent $a_v$.
 For each $b\in\{\mathsf T,\mathsf F\}$, let
 \[
 p_{a_v,J_v^b}=1,\qquad c_{a_v,J_v^b}=1/2.
 \tag{3}\label{eq:var-gadget}
 \]
 Thus each of the two tasks gives agent $a_v$ welfare $1/2$,
 and \eqref{ef_cons_eq} requires $\alpha_{J_v^b}\ge 1/2$ if agent $a_v$
 is assigned $J_v^b$. Since $a_v$ can receive at most one task,
 the selected task will encode one Boolean value for the variable $x_v$.

 For a literal occurrence $\ell$ involving $x_v$, define its
 \emph{falsifying task} by
 \[
 J(\ell)=
 \begin{cases}
 J_v^{\mathsf F},&\ell=x_v,\\
 J_v^{\mathsf T},&\ell=\neg x_v.
 \end{cases}
 \tag{4}\label{eq:falsifying-root}
 \]
 Hence $J(\ell)$ is the task chosen by the variable agent precisely
 when $\ell$ is false under the encoded assignment.

For each clause $C_t$ with its three literal occurrences
 $\ell_{t,1},\ell_{t,2},\ell_{t,3}$, introduce a common task $G_t$
 with reward $2/R$, private tasks $P_{t,h}$ of reward $1/R$
 for $h\in\{1,2,3\}$, three literal agents $b_{t,h}$, and a
 dummy agent $d_t$. Specify the following positive-probability pairs:
 \[
 \begin{array}{c|c|c|c}
 \text{agent}&\text{task}&p_{i,j}&c_{i,j}\\ \hline
 b_{t,h}&P_{t,h}&1&0\\
 b_{t,h}&G_t&1&0\\
 b_{t,h}&J(\ell_{t,h})&\delta/R&0\\
 d_t&G_t&\eta/2&0
 \end{array}
 \tag{5}\label{eq:clause-gadget}
 \]
 The corresponding welfare values are $1/R$, $2/R$, $\delta/R$, and
 $\eta/R$, respectively. In particular, a literal agent has
 exactly three tasks of interest: its private task, the common
 clause task, and the falsifying task. Every variable agent has
 two interests and every dummy agent has one. Hence, it is a $3$-interest instance.

\paragraph{Upper bounds and the threshold.}
A root task $J_v^b$ can provide welfare $1/2$ to its
 variable agent, welfare $\delta/R$ to a literal agent, or zero
 to any other agent. Since $a_v$ can take at most one root
 and each root is assigned at most once, the total welfare
 from both root tasks of $x_v$ is at most
 \[
 \begin{cases}
 \frac12+\frac\delta R,&\text{if }a_v\text{ receives a root},\\
 \frac{2\delta}R,&\text{if }a_v\text{ receives no root}.
 \end{cases}
 \tag{6}\label{eq:root-bound}
 \]
 This remains true if multiple literal occurrences point
 to the same root task, because that task has unit capacity.

 Within clause $C_t$, the three private tasks have welfare
 at most $1/R$ each and $G_t$ has welfare $2/R$ when assigned
 to a literal agent and $\eta/R$ when assigned to the dummy agent.
 If $G_t$ is assigned to a literal agent, that same literal
 agent cannot also take its private task; hence the total
 welfare from tasks $G_t,P_{t,1},P_{t,2},P_{t,3}$ is at most
 \[
 \frac2R+\frac1R+\frac1R=\frac4R.
 \tag{7}\label{eq:clause-bound}
 \]
 If $G_t$ is \emph{not} assigned to a literal agent, the
 total is at most $(3+\eta)/R$. These bounds apply regardless
 of how any root tasks are assigned.

 Suppose that a feasible solution has $\texttt{Wel}(S, \alpha)\ge K$.
 If some variable agent does not receive a root, then by
 \eqref{eq:root-bound} and \eqref{eq:clause-bound},
 \[
 \texttt{Wel}(S, \alpha) \le (n-1)(\frac{1}{2}+\frac{\delta}{R}) + \frac{2\delta}{R} + m\frac{4}{R}
 \le\frac{n-1}{2}+\frac{2n\delta}{R}+\frac{4m}{R}
 <\frac n2+\frac{4m}{R}=K,
 \]
 contradicting assumption.
 Thus \emph{every variable agent receives exactly one root task}.

 If some clause task $G_t$ is not assigned to one of its three
 literal agents, the bounds above now give
 \[
 \texttt{Wel}(S, \alpha)
 \le\frac n2+\frac{n\delta}{R}+\frac{4m-1+\eta}{R}
 <K,
 \]
 again contradicting assumption.
 Consequently, \emph{every clause task $G_t$ is assigned
 to a literal agent} in any feasible outcome attaining $K$.
 These are the two structural facts needed for soundness.

\smallskip\noindent
\textbf{Completeness: satisfiability implies welfare at least $K$.}
Assume $\varphi$ is satisfied by a truth assignment $\sigma$.
 For every variable $x_v$, assign task $J_v^{\sigma(x_v)}$
 to $a_v$ and give this root task the share $\alpha_{J_v^{\sigma(x_v)}} = 1/2$.
 For each clause $C_t$, choose one occurrence $\ell_{t,h_t}$
 that is true under $\sigma$. Assign $G_t$ to agent $b_{t,h_t}$,
 set $\alpha_{G_t}=0$, and assign each of the two other
 literal agents its private task $P_{t,h}$ with share
 $\alpha_{P_{t,h}}=1$. Leave all dummy agents, the selected
 literal's private task, and the unchosen variable root tasks
 unassigned. Set contracts for all other tasks to zero.

 We verify all effort and EF constraints, including those of
 unmatched agents. Each variable agent receives utility
 $\tfrac12-1/2=0$ and is interested only in its two roots;
 the other root is unassigned. The literal agent receiving
 $G_t$ has utility zero and its falsifying root is unassigned,
 because the chosen literal is true. Its private task is
 also unassigned. Every other literal agent has utility $1/R$
 from its private task. It values the assigned common task
 at $(2/R)\alpha_{G_t}=0$ and any assigned falsifying root at
 most $\delta/(2R)<1/R$, so it does not envy either.
 Finally, each unmatched dummy agent values only $G_t$,
 which yields utility $(\eta/R)\alpha_{G_t}=0$.
 Hence the constructed outcome satisfies \eqref{ef_cons_eq} and \eqref{eff_constraint}.
 Its welfare is exactly
 \[
 \frac n2
 +\sum_{t=1}^m\left(\frac2R+\frac1R+\frac1R\right)
 =\frac n2+\frac{4m}{R}=K.
 \]

\smallskip\noindent
\textbf{Soundness: welfare at least $K$ implies satisfiability.}
Let $(S,\alpha)$ be any feasible outcome with
 $\texttt{Wel}(S, \alpha)\ge K$. By the upper-bound argument, every
 variable agent $a_v$ receives one of its two root tasks, by which we assign each variable a Boolean value and have a truth assignment $\sigma$.
 Moreover, the upper-bound argument implies that every $G_t$
 is assigned to some literal agent $b_{t,h_t}$.

 Since dummy task $d_t$ values only $G_t$ and cannot receive it,
 it has zero utility, whether it is unmatched or has been
 assigned a zero-value task. Its EF constraint against
 $G_t$ yields
 \[
 0\ge U_{d_t,G_t}(\alpha_{G_t})
 =\frac{\alpha_{G_t}\eta}{R}.
 \]
 Since $\eta>0$ and $\alpha_{G_t}\ge0$, necessarily
 \[
 \alpha_{G_t}=0.
 \tag{8}\label{eq:zero-clause-share}
 \]
 Thus $b_{t,h_t}$ receives utility
 $U_{b_{t,h_t},G_t}(\alpha_{G_t})=0$.
 Suppose the chosen literal $\ell_{t,h_t}$ were false under
 $\sigma$: (1) If $\ell_{t,h_t}=x_v$, then its falsifying task $J(\ell_{t,h_t}) = J_v^F$ and agent $a_v$ is also assigned task $J_v^F$ since literal $\ell_{t,h_t} = x_v$ were false; (2) If $\ell_{t,h_t}=\neg x_v$, then its falsifying task $J(\ell_{t,h_t}) = J_v^T$ and agent $a_v$ is also assigned task $J_v^T$ by our construction;
 Hence, 
 \eqref{eff_constraint} for that variable agent and \eqref{eq:var-gadget}
 imply
 \[
 \alpha_{J(\ell_{t,h_t})}\ge\frac12.
 \]
 But then the literal agent would obtain from this
 \emph{assigned} falsifying task utility at least
 \[
 U_{b_{t,h_t},J(\ell_{t,h_t})}
       (\alpha_{J(\ell_{t,h_t})})
 =\frac\delta R\alpha_{J(\ell_{t,h_t})}
 \ge\frac{\delta}{2R}>0,
 \]
 strictly greater than its utility from $G_t$.
 This violates EF. Therefore the value of literal $\ell_{t,h_t}$ is true.
 Since this holds for every clause, $\sigma$ satisfies
 $\varphi$.

\medskip
There are $2n+4m$ tasks and $n+4m$ agents, and all
 numerators and denominators in \eqref{eq:parameters}--
 \eqref{eq:clause-gadget} (including $1/R$ and $2/R$)
 have polynomial bit length.
 The construction is therefore computable in polynomial time.
 This concludes the proof.
\end{proof}

\section{Connection to Envy-Free Pricing}\label{sec:pricing}
In this section, we relate our fair contract design problem to the well-studied envy-free pricing problem in the literature \citep{anshelevich2017envy,guruswami2005profit}. 

In addition to welfare maximization in \eqref{program-welfare}, we consider maximizing the principal's expected revenue. For a unit-demand allocation represented by a matching $M$, define
\begin{equation}\label{eq:contract_revenue}
 \operatorname{Rev}(M,\alpha)
 = \sum_{i \in \mathcal{A}} \sum_{j\in S_i}(1-\alpha_j)p_{i,j}r_j.
\end{equation}
Unlike welfare, this objective depends on the contracts. 
As discussed in \citet{castiglioni2025algorithmic}, revenue maximization in the envy-free pricing problem is straightforward in the additive valuation settings: the optimal price for each item is set as the maximum value of agents. In contrast, the fair contracts problem (with the objective of maximizing revenue) is APX-hard even in additive cases.

In particular, we establish a connection to the unit-demand, unit-supply envy-free pricing problem~\citep{chen2010envy}: There are $m$ heterogeneous items and $n$ buyers. The seller sets prices $(p_1, p_2, \dots, p_m)$ for $m$ items, and each buyer chooses its favorite item subject to non-negative utility constraints. The buyer $i\in\mathcal{A}$ has valuation $v_{i,j}\ge 0$ in the item $j\in\mathcal{T}$. A buyer assigned item $j$ has utility $v_{i,j}-q_j$, and an unmatched buyer has utility zero. Feasibility requires every buyer's utility to be nonnegative and at least $v_{i,t}-q_t$ for \emph{every} item $t$, including unsold items. 


\begin{assumption}\label{assump_epdn_c1}
There is a common success probability $p\in[0,1]$ such that $p_{i,j}=p$ for every agent--task pair, and $r_j=r$, where $r\ge 0$ for every task.
\end{assumption}

We continue to consider the unit-demand agents. The next proposition shows that under \cref{assump_epdn_c1}, the revenue-maximizing problem in fair contracts is equivalent to the envy-free pricing problem.

\begin{proposition}\label{reudction_eachotuerh}
    Consider unit-demand agents. Under Assumption \ref{assump_epdn_c1}, the fair contracts problem with revenue (resp. welfare) maximization is equivalent to the unit-demand unit-supply envy-free pricing problem with revenue (resp. welfare) maximization. In other words, there exists a polynomial-time reduction such that these two problems reduce to each other, where the principal and seller obtain the same revenue (resp. welfare).
\end{proposition}
\begin{proof}

Rewrite the utility of agent $i$ exerting efforts on task $j$ under contract $\alpha_j$ as follows,
\begin{equation}\label{eq:pricing_utility_identity}
 U_{i,j}(\alpha_j)=W_{i,j}-(1-\alpha_j)pr,
 \qquad W_{i,j}=pr-c_{i,j}.
\end{equation}
Thus $W_{i,j}$ acts as a valuation and $(1-\alpha_j)pr$ as a price in envy-free pricing problems.

    \paragraph{From pricing to contracts.} Consider any feasible prices $(p_1, p_2, \dots, p_m)$ and allocations $(s_1, s_2, \dots, s_n)$ in envy-free pricing. If $s_i =0 $, it means that agent $i$ is not assigned a task.  Without loss of optimality, we assume prices $\max~ \{p_i\} \le \max~ \{v_{i, j}\}$. Constructing a contract design instance $p>0, r$ and $c_{i, j} \ge 0, \forall i, j$ such that $pr - c_{i, j} =v_{i, j}$. 
    Set the contract for task $j$ as $\alpha_j = 1- \frac{p_j}{pr}$. Clearly, $\alpha_j \in [0, 1]$ by construction. The \eqref{ef_cons_eq} holds since
    \[
    \alpha_{s_i} pr - c_{i, s_i} =  pr - c_{i, s_i} - p_{s_i} =  v_{i,s_i} - p_{s_i} \ge   \max\{0, v_{i, s_k} - p_{s_k} \} =  \max\{0, \alpha_{s_k} pr - c_{i, s_k} \} 
    \]
    where the inequality is due to that each agent $i$ chooses its favorite item.
 The revenue in fair contracts is the same as that in envy-free pricing: $\operatorname{Rev}(M,\alpha)
 = \sum_{i \in \mathcal{A} \& s_i \neq 0 }(1-\alpha_{s_i})pr= \sum_{i \in \mathcal{A} \& s_i \neq 0 } p_{s_i}$. Moreover, their welfare is also the same: $\sum_{i \in \mathcal{A} \& s_i \neq 0 }W_{i, j}= \sum_{i \in \mathcal{A} \& s_i \neq 0 } v_{i,j}$.

\paragraph{From contracts to pricing.} 
Consider any feasible contracts $(\alpha_1, \alpha_2, \dots, \alpha_m)$ and allocation of tasks $(s_1, s_2, \dots, s_n)$ in fair contracts, where the contracts of unassigned tasks are set to zero without loss of optimality.  If $s_i =0$, it means that agent $i$ is not assigned a task. 
We can set the price for task $j$ as $p_{j} = (1-\alpha_{j})pr$, and set $v_{i, j} =\max\{ W_{i, j}, 0\}$ as the agent $i$'s valuation for item $j$. 
We only need to verify that $s_i$ is the favorite item for agent $i$ in envy-free pricing. 
Obviously, $s_i$ gives weakly higher utility to agent $i$ than any other task $s_j$, i.e., 
\begin{align*}
   v_{i, s_i} - p_{s_i} &= W_{i, s_i} - (1-\alpha_{s_i})pr\\
   &\ge \max\{ 0, W_{i, s_k} - (1-\alpha_{s_k})pr\} \\
   &= \max\{ 0, W_{i, s_k} - (1-\alpha_{s_k})pr\} = \max\{ 0, W_{i, s_k} - p_{s_k}\} \\
   &= \max\{0, v_{i,s_k} - p_{s_k}\}
\end{align*}
where the inequality is due to \eqref{ef_cons_eq}. The two instances give the principal the same revenue: $\sum_{i \in \mathcal{A} \& s_i \neq 0 } p_{s_i} = \sum_{i \in \mathcal{A} \& s_i \neq 0 }(1-\alpha_{s_i})pr$. Moreover, their welfare is also the same: $\sum_{i \in \mathcal{A} \& s_i \neq 0 } v_{i,j} = \sum_{i \in \mathcal{A} \& s_i \neq 0 }W_{i, j}$.
    
Therefore, the two problems can be reduced to each other in polynomial time, while preserving the principal/seller's revenue. Hence, these two problems are equivalent.
\end{proof}

By the above equivalence, we have the following two straightforward corollaries. 


The first corollary is that for settings of unit-demand agents, maximizing the principal's expected revenue in fair contracts is $\NP$-hard. The following result is immediate by the equivalence and the results in \citep{chen2010envy}: revenue maximization in unit-demand, unit-supply envy-free pricing is $\NP$-hard~\citep{chen2010envy}. 

\begin{corollary}\label{cor:revenue_hardness}
For unit-demand agents, maximizing the principal's expected revenue subject to the effort and EF constraints is $\NP$-hard, even when the success probabilities for different task-agent pairs the same (i.e., Assumption~\ref{assump_epdn_c1}).
\end{corollary}

The second corollary is due to the fact that the welfare is the same in two problems. Recall that a Walrasian equilibrium maximizes the social welfare~\citep{blumrosen2007combinatorial}. Moreover, the Walrasian equilibrium can be found in polynomial time by computing the maximum weight matching in the bipartite graph. 
This implies that in the unit-demand unit-supply envy-free pricing problem, the welfare-maximizing solution can be found in polynomial time, which immediately implies the following corollary by the equivalence.

Second, unit-demand, unit-supply pricing instances admit a Walrasian equilibrium: an envy-free outcome with nonnegative prices in which every unsold item has price zero. It is known that such an equilibrium maximizes social welfare~\citep{blumrosen2007combinatorial}. Its allocation and prices can be computed in polynomial time using a maximum-weight bipartite matching and an optimal dual solution to the matching linear program. Consequently, welfare maximization in envy-free pricing is polynomial-time solvable, and the equivalence gives the following corollary.

\begin{corollary}\label{cor:welfare_poly}
Consider unit-demand agents. Under \cref{assump_epdn_c1}, a welfare-maximizing solution subject to the effort and EF constraints can be found in polynomial time.
\end{corollary}

\section*{Disclosure of AI Use}
The authors used GPT-6.1 Sol for language polishing, manuscript drafting, and mathematical reasoning assistance. The authors verified the correctness and originality of all content, including references. 

The authors conceived and formulated all research problems studied in this paper.

GPT was used to help draft the abstract and \cref{section_intro}, and to develop proofs of the hardness results in \cref{sec:expanded-welfare-hardness}. These drafts and proofs were subsequently reviewed and revised by the authors. The results in \cref{hardness_general_case_sec} and \cref{sec:pricing} were developed by the authors.
For \cref{polynotimaitmelaoghtm}, the authors developed the core proof strategy and identified the key tree structures, drawing inspiration from \citet{chen2010envy}. The structural characterization of minimum matchings and the proofs of their properties were developed primarily by the authors, with assistance from AI. The proof of graph perfectness was developed by AI. AI was also used to correct errors. 

\bibliographystyle{apalike}
\bibliography{refer}
\end{document}